\documentclass[sigconf,screen,authorversion]{acmart}
 
\AtBeginDocument{%
  \providecommand\BibTeX{{%
    \normalfont B\kern-0.5em{\scshape i\kern-0.25em b}\kern-0.8em\TeX}}}

\copyrightyear{2025}
\acmYear{2025}
\setcopyright{acmlicensed}
\acmConference[CHI '25]{CHI Conference on Human Factors in Computing Systems}{April 26--May 1, 2025}{Yokohama, Japan}
\acmBooktitle{CHI Conference on Human Factors in Computing Systems (CHI '25), April 26-May 1, 2025, Yokohama, Japan} 
\acmPrice{15.00}
\acmDOI{10.1145/3706598.3714257}
\acmISBN{979-8-4007-1394-1/25/04}

\usepackage{colortbl}
\usepackage{xcolor}

\usepackage{acmart-taps}
\usepackage{enumitem}
\usepackage{graphicx}
\usepackage{soul}
\usepackage{multirow, multicol}
\usepackage{wrapfig}
\definecolor{tablerowcolor}{HTML}{999999}
\definecolor{lighttablerowcolor}{HTML}{b2b2b2}
\definecolor{color1}{RGB}{229, 148, 12}
\definecolor{color2}{RGB}{56, 152, 171}
\definecolor{color3}{RGB}{166,86,40}
\definecolor{RRcha}{HTML}{000000} %{9bbdef}

\newcommand{\etal}{et~al.}

\newcommand{\q}[1]{\textit{``#1''}}
\newcommand{\qq}[1]{{``#1''}}
\def\signed #1{{\leavevmode\unskip\nobreak\hfil\penalty50\hskip2em
  \hbox{}\nobreak\hfil(#1)%
  \parfillskip=0pt \finalhyphendemerits=0 \endgraf}}

\newsavebox\mybox

\definecolor{BLUE}{rgb}{0,0,1}
\definecolor{BLACK}{rgb}{0,0,0}
\newcommand{\RR}[1]{\textcolor{black}{#1}}
\newcommand{\CR}[1]{\textcolor{black}{#1}} % copyright

\newcommand{\wrap}[2][1.5cm]{
  \begin{wrapfigure}{l}{\dimexpr #1 - 0.65cm \relax}
    \centering
    \vspace{-.25cm}
    \includegraphics[width=#1]{#2}
           \vspace{-0.8cm}
  \end{wrapfigure}
}

\begin{document}


\title[``You'll Be Alice Adventuring in Wonderland!'' Processes, Challenges, and Opportunities of Creating Animated VR Stories]{``You'll Be Alice Adventuring in Wonderland!'' Processes, Challenges, and Opportunities of Creating Animated Virtual Reality Stories}

\author{Lin-Ping Yuan}
\affiliation{%
  \institution{The Hong Kong University of Science and Technology}
  \city{Hong Kong SAR}
  \country{China}}
\email{yuanlp@cse.ust.hk}
\orcid{0000-0001-6268-1583}

\author{Feilin Han}
\affiliation{%
  \institution{Beijing Film Academy}
  \city{Beijing}
  \country{China}}
\email{hanfeilin@bfa.edu.cn}
\orcid{0000-0001-7463-2252}

\author{Liwenhan Xie}
\affiliation{%
  \institution{The Hong Kong University of Science and Technology}
  \city{Hong Kong SAR}
  \country{China}}
\email{liwenhan.xie@connect.ust.hk}
\orcid{0000-0002-2601-6313}

\author{Junjie Zhang}
\affiliation{%
  \institution{The Hong Kong University of Science and Technology (Guangzhou)}
  \city{Guangzhou}
  \country{China}
  }
\affiliation{%
  \institution{The Hong Kong University of Science and Technology}
  \city{Hong Kong SAR}
  \country{China}}  
\email{jakezhang@hkust-gz.edu.cn}
\orcid{0000-0002-3155-5805}

\author{Jian Zhao}
\affiliation{%
  \institution{University of Waterloo}
  \city{Waterloo, ON}
  \country{Canada}}
\email{jianzhao@uwaterloo.ca}
\orcid{0000-0001-5008-4319}

\author{Huamin Qu}
\affiliation{%
  \institution{The Hong Kong University of Science and Technology}
  \city{Hong Kong SAR}
  \country{China}}
\email{huamin@cse.ust.hk}
\orcid{0000-0002-3344-9694}

% \renewcommand{\shortauthors}{L. Yuan, F. Han, L. Xie, J. Zhang, J. Zhao, and H. Qu}

\begin{abstract}
  Animated virtual reality (VR) stories, combining the presence of VR and the artistry of computer animation, offer a compelling way to deliver messages and evoke emotions. Motivated by the growing demand for immersive narrative experiences, more creators are creating animated VR stories. However, a holistic understanding of their creation processes and challenges involved in crafting these stories is still limited. Based on semi-structured interviews with 21 animated VR story creators, we identify ten common stages in their end-to-end creation processes, ranging from idea generation to evaluation, which form diverse workflows that are story-driven or visual-driven. Additionally, we highlight nine unique issues that arise during the creation process, such as a lack of reference material for multi-element plots, the absence of specific functionalities for story integration, and inadequate support for audience evaluation. We compare the creation of animated VR stories to general XR applications and distill several future research opportunities.
\end{abstract}




% \keywords{datasets, neural networks, gaze detection, text tagging}




\maketitle
\section{Introduction}
An animated virtual reality (VR) story is a sequence of connected events, crafted to convey messages and evoke emotions, portrayed through computer animation, and unfolded within an interactive and three-dimensional VR environment~\cite{franklin2021fairy,cutler2019making,dooley2021cinematic, bosworth2018craftingVRstory}. 
Combining the high level of presence offered by VR~\cite{bindman2018bunny} with the boundless creative possibilities of computer animation~\cite{cutler2019making}, animated VR stories can provide unparalleled experiences.
These advantages have attracted many creators to tell their stories in this form. 
Early pioneers included filmmakers, animators, and transmedia artists from Oculus Story Studio and Google's Spotlight Stories, who created groundbreaking shorts like \textit{Henry} (2015) and \textit{Pearl} (2016). Building on this momentum, animated VR stories have now become mainstream within the VR sections~\cite{bosworth2018craftingVRstory} of prominent film festivals \cite{CannesAward} like Cannes and Sundance, as evidenced by award-winning stories \textit{Invasion!} (2017), \textit{The Dream Collector} (2017), \textit{Baba Yaga} (2020), and \textit{Namoo} (2022). 
Recently, with an expanding VR consumer market and a growing demand for VR content, more and more creators are releasing their animated stories~\cite{CannesAward} as standalone VR programs or on platforms like Oculus Video Animation Players.

\begin{figure*}[!tb]
    \centering
    \includegraphics[width=\linewidth]{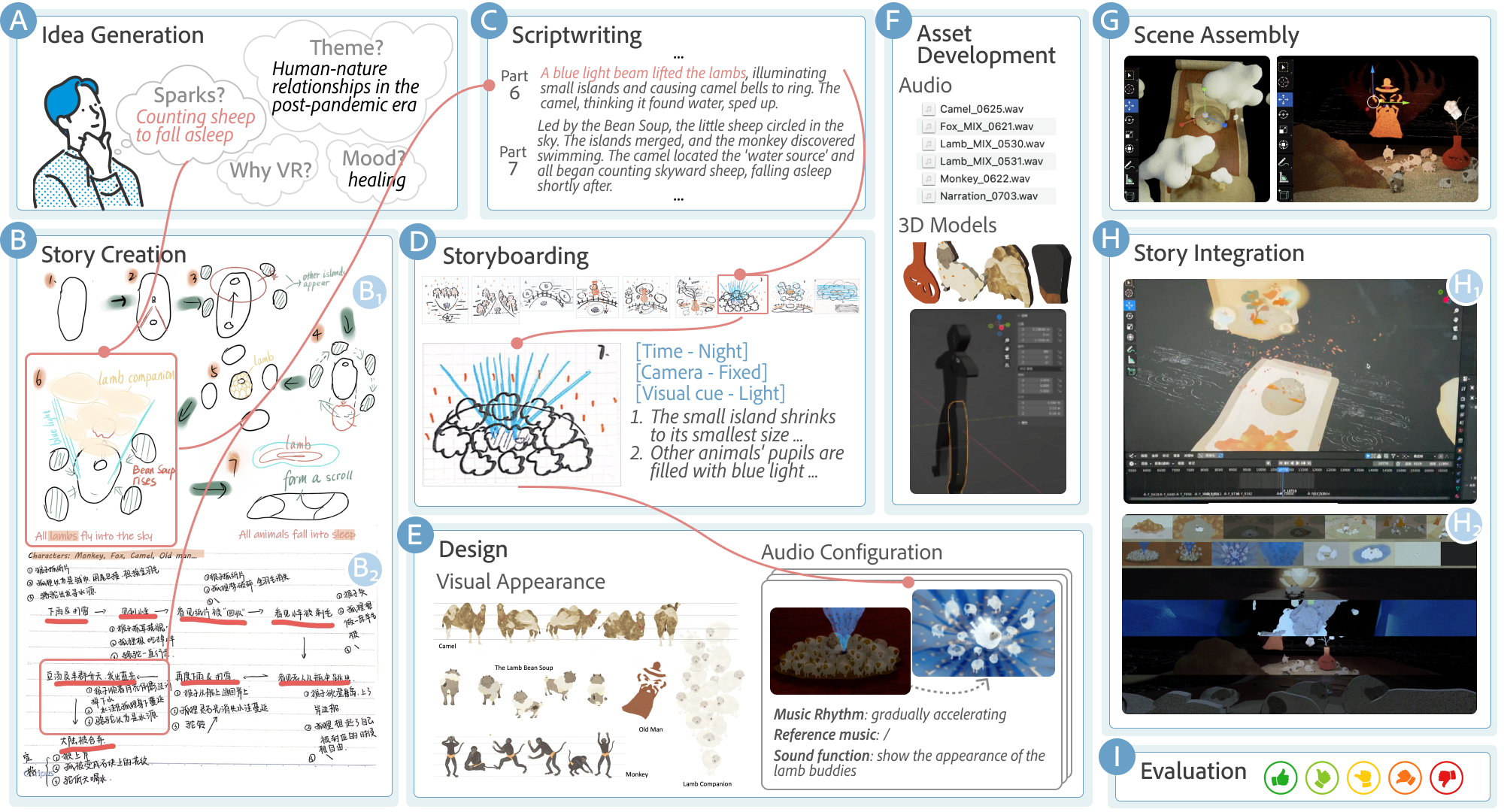}
    \caption{An illustration of 9 stages in creating an animated VR story (\CR{\copyright Coin's team}), where the red lines indicate how a storyline evolves.
    (A) \textbf{Idea generation}: brainstorming general ideas. (B) \textbf{Story creation}: conceiving the main character's activities using sketches (B1) and developing the story by including other characters with scattered word pieces (B2). (C) \textbf{Scriptwriting}: transforming the stories into textual scripts. (D) \textbf{Storyboarding}: sketching the key moments while configuring the timing, camera setups \& visual cues. (E) \textbf{Design}: specifying visual (e.g., character appearance) \& auditory (e.g., music) elements. (F) \textbf{Asset development}: developing rigged and animated 3D models \& audio clips. (G) \textbf{Scene assembly}: assembling individual assets spatially into scenes in the story. (H) \textbf{Story integration}: aligning multiple story elements on the timeline and integrating the whole story. (I) \textbf{Evaluation}: collecting feedback from viewers.} 
    \label{fig:workflow_example_1}
    \Description{This figure showcases the nine stages (labeled with Part A-I). Part A ("Idea Generation") shows a person thinking about various themes, such as "Counting sheep to fall asleep" and "Why VR?". Part B ("Story Creation") contains a rough sketch and text snippets illustrating the main character's activities, such as animals flying into the sky. Part C ("Scriptwriting") provides a script sample, describing how "a blue light beam lifted the lambs". Part D ("Storyboarding") presents a sequence of small visual frames, detailing information such as "Time: Night”, "Camera: Fixed”, and "Visual cue: Light”. Part E ("Design") features concepts for the visual appearance of various characters, such as a camel and a monkey. Part F ("Asset Development") focuses on audio and 3D models. It lists sound files and displays basic 3D model designs, including textured models of the characters. Part G ("Scene Assembly") contains two screenshots of a 3D environment being built, showing the placement of various elements in a virtual scene. Part H ("Story Integration") shows two additional 3D-rendered images, illustrating how the scenes are temporally connected. Part I ("Evaluation") shows a set of thumbs-up/thumbs-down buttons to indicate feedback or assessment of the final outcome.}
\end{figure*}

Despite the increasing interest in animated VR stories, dedicated support for creators has not received adequate attention from industry or academia.
For example, commercial software either lacks specialized design for animated VR stories (e.g., Blender, Unity) or is no longer actively maintained (e.g., Quill).
Although researchers have proposed several authoring tools~\cite{galvane2019vr, stemasov2023sampling, wang2022videoposevr, nguyen2017collavr, vogel2018animationvr}, these tools are fragmented and limited to a subset of tasks like VR storyboards~\cite{henrikson2016multi, henrikson2016storeoboard, galvane2019vr} and asset animation~\cite{lamberti2020immersive, vogel2018animationvr, wang2022videoposevr}. Many other essential tasks reported by creators~\cite{ward2021tinker,gipson2018disneycicles,cutler2019making, darnell2016invasion} are not adequately supported, such as composing VR scenes~\cite{cutler2019making,gipson2018disneycicles} and enabling interactions inside stories~\cite{ward2021tinker,darnell2016invasion}.
The insufficient support compels VR story creators to bridge general-purpose software, demanding significant time and effort to learn and experiment~\cite{gipson2018disneycicles}. 
They wish for streamlined tools that blend into their processes and address the challenges they face~\cite{cutler2019making}.
However, a comprehensive understanding of current creation processes and challenges remains absent, making the future research directions in supporting animated VR story creation unclear.

Recent HCI studies~\cite{ashtari2020creating, krauss2021current,  krauss2022elements} have provided empirical insights into the current practices and challenges of crafting VR applications. Although they acknowledge the complexities of prototyping story-driven VR experiences~\cite{ashtari2020creating, krauss2021current}, these studies generally aim to accommodate a wide array of VR use cases such as training and rehabilitation. 
Consequently, creators' unique considerations for animated VR stories have been overlooked.
The first consideration is rooted in the essence of stories.
Animated VR stories prioritize aspects such as narrative engagement, empathy, and emotional resonance~\cite{bindman2018bunny, bahng2020reflexive}. To achieve so, creators carefully consider a blend of visual, auditory, and interactive story elements~\cite{cutler2019making,ward2021tinker} and may face challenges in orchestrating them cohesively in both spatial and temporal aspects.
The second consideration is rooted in the use of VR and computer animation technologies to tell the story. 
Creators need to consider the benefits and constraints these technologies present. For example, though VR enables storytelling with a high level of presence, it also requires appealing 360-degree visuals~\cite{cutler2019making,gipson2018disneycicles}, which may bring challenges to balance visual quality and runtime performance.
We have yet to adequately understand how the interplay of the two considerations poses challenges to animated VR story creators within their creation processes.

To fill the above gaps, this study aims to answer the following research questions:

\begin{itemize}[left=0pt]
    \item \textbf{RQ1}: What are the creation processes that creators usually follow when crafting animated VR stories?
    \item \textbf{RQ2}: What challenges do animated VR creators face, especially when marrying story elements with the benefits and constraints of VR and computer animation technologies?
\end{itemize}

We conducted semi-structured interviews with 21 animated VR story creators. To achieve a broad understanding, we ensured our interviewees' backgrounds covered experiences in crafting diverse stories. The stories varied in their visual styles and levels of interactivity, from head-tracking to gesture interactions with in-story characters. They were made with various non-immersive (e.g., Blender, Unity) or immersive (e.g., Quill, Open Brush) software.
%
% findings
For RQ1, we identify ten common stages in the creation processes, from idea generation to evaluation (Fig.~\ref{fig:workflow_example_1}). The inclusion and order of these stages can vary and form diverse workflows. Two types of workflows emerge: story-driven and visual-driven workflows that prioritize story content or visuals, respectively. 
For RQ2, we identify seven challenges including a total of seventeen issues (Table~\ref{tab:challenges}) at different stages. Among them, nine issues are relatively unique to animated VR stories, which highlight narrative intent and viewer autonomy, satisfactory visuals for artistic expression, and multiple narrative perspectives. 
Based on the findings, we offer several future research opportunities to support animated VR story creation and discuss the differences between animated VR stories and general XR applications~\cite{krauss2021current, ashtari2020creating, krauss2022elements}. 

In summary, our contributions around animated VR stories are threefold: (1) identification of ten common stages and two types of workflows in the creation processes (\RR{Sec.~\ref{sec:processes}}), (2) summarization of nine unique issues in crafting them (\RR{Sec.~\ref{sec:challenges}}), and (3) provision of future research opportunities to support their creation (\RR{Sec.~\ref{sec:opportunities}}).
\section{Background and Related Work}

Our study builds on existing research into storytelling guidelines and design considerations for animated VR stories. To contextualize our research outcomes within HCI literature, we examine empirical studies on the creation of general XR applications and authoring tools for animated VR stories.

\subsection{Guidelines for Animated VR Stories}
Animated VR stories differ from traditional animations in two main characteristics~\cite{cutler2019making, godde2018cinematic}. First, they are set within immersive 3D spaces that encircle viewers~\cite{godde2018cinematic, serrano2017movie}, rather than being projected onto a flat screen. Second, they offer viewers various levels of interactivity~\cite{tong2021viewer, rothe2019interactioninCVR}, such as turning heads, walking around, and directly interacting with characters. 
These characteristics reconfigure the conventional relationships between the audience, camera, and story~\cite{henrikson2016multi, tong2022adaptive}. 
In contrast to traditional animations where creators fully control the camera and story, in VR, viewers pilot the cameras and can actively participate in and influence the story. 
As a result, some typical storytelling techniques, such as frame control and camera movement~\cite{serrano2017movie,henrikson2016multi}, may not be applicable to VR. 
Therefore, practitioners and researchers have proposed tailored storytelling guidelines~\cite{cutler2019making, williams2021virtual, aitamurto2021fomo, gupta2020roleplaying}, such as guiding user attention through audiovisual cues~\cite{rothe2019guidance, schmitz2020directing}, and strategically distributing story elements across 3D spaces~\cite{kvisgaard2019frames, pope2017geometry}.

HCI research has further broadened the research scope by associating VR storytelling with user experiences (e.g., presence~\cite{bindman2018bunny, kroma2022technical, men2017impact}, narration comprehension~\cite{bindman2018bunny, gupta2020investigating}, motion sickness~\cite{han2022evaluating, rahimi2018scene}, embodiment~\cite{liu2022generating}), emotional responses (e.g., empathy~\cite{bindman2018bunny}, affect~\cite{norouzi2021virtual}), and cognitive processes (e.g., self-reflection~\cite{bahng2020reflexive}, knowledge acquisition~\cite{zhang2019exploring, hwang2022being, lee2020data}, situational awareness~\cite{zhu2024reader}), as well as providing design recommendations based on study results. 
For instance, Bindman~\etal~\cite{bindman2018bunny} discovered that viewers' narrative engagement and empathy were more influenced by their perceived roles within the story rather than by the level of device immersion. 
Bahng~\etal~\cite{bahng2020reflexive} created an interactive story about death and loneliness, identified four reflexive design factors, and suggested incorporating these reflexive factors in future VR storytelling experiences, particularly when the goal is to provoke thoughtful self and social reflection.

While these studies alleviate creative hurdles, they neglect the practical difficulties of execution. Our study builds on these guidelines, examining how creators consider them during the creation process and identifying barriers to applying them in their stories.

\subsection{Empirical Studies for General XR Creation}
Recent HCI research has explored the current practices, challenges, and opportunities associated with creating XR applications. These studies target various types of creators (e.g., hobbyists~\cite{ashtari2020creating}, professional developers~\cite{krauss2021current, liu2023challenges, borsting2022software}, and professional designers~\cite{ashtari2020creating, krauss2021current, krauss2022elements}), settings (e.g., industry~\cite{krauss2022elements} and non-industry~\cite{ashtari2020creating, shin2023space}), and phases (e.g., the whole process~\cite{ashtari2020creating, krauss2021current, borsting2022software} and testing phase~\cite{liu2023challenges}). 

Some of these studies~\cite{ashtari2020creating, krauss2021current, borsting2022software, krauss2022elements, mendez2025immersivesurvey} recognize the significant difficulties in immersive storytelling. 
For example, Ashtari~\etal~\cite{ashtari2020creating} reported the difficulty in designing an immersive story with real-world sensory experiences and engagement, although VR provides immersive environments with reduced real-world distractions. 
Krauß~\etal~\cite{krauss2022elements} noted that typical manifestations of prototypes like text and storyboards fell short in effectively conveying the feeling of XR. However, these studies primarily focus on general XR applications and lack deeper insights into the unique characteristics of animated VR stories. Furthermore, their interviewees did not fully cover the current main creators of animated VR stories, who are probably filmmakers, transmedia artists, and animators~\cite{henrikson2016multi,rall2022pericles, cutler2019making}. 
Shin and Woo~\cite{shin2023space} found that creators adopted different creative strategies to associate story events with physical landmarks in AR. However, these findings may not be applicable to animated VR stories, which take place in virtual worlds and whose storylines are not constrained by physical sites.

Our study complements this line of research by identifying the creation processes and challenges of animated VR stories. 
Since creators need to consider multiple story elements, as well as benefits and constraints brought by VR and computer animation technologies, we are particularly interested in how these considerations influence their processes and pose challenges. 
Based on our findings, we compare animated VR stories to general XR applications.

\begin{table*}[t]
    %\scriptsize
    \setlength{\aboverulesep}{0.2pt}
    \setlength{\belowrulesep}{0.2pt}
\setlength{\tabcolsep}{2pt}
\centering
\caption{A summary of interviewees' background and relevant experience. From left to right, each column shows the interview ID, background, years (Y) of experiences in general art/design and VR, number (\#) of animated VR stories and all VR artworks being created, and a list of different types of non-VR artworks.}
\label{tab:participant_information}
\Description{A summary of participants’ background and relevant experience.}
\resizebox{\textwidth}{!}{\begin{tabular}{llccccl}
 \toprule
 \multirow{2}[0]{*}{\textbf{ID}}  & \multicolumn{1}{c}{\multirow{2}[0]{*}{\textbf{Background}}} &  \multicolumn{2}{c}{\textbf{Experience} (Y)} & \multicolumn{2}{c}{\textbf{VR Artworks } (\#)} & \multicolumn{1}{c}{\multirow{2}[0]{*}{\textbf{Experience in Non-VR Artworks} (Types) } } \\ \cline{3-6}
 & & \multicolumn{1}{c}{\textbf{Art}}  & \multicolumn{1}{c}{\textbf{VR}} & \multicolumn{1}{c}{\textbf{Story}} & \multicolumn{1}{c}{\textbf{All}} &  \\ \midrule                                                               
P1          & Digital Media Arts                 & 10                             & 4           & 2                          & 2              & Painting, 2D/3D Animation                                                           \\
P2          & Transmedia Arts                    & 10                             & 2           & 2                          & 2              & Comics,  Painting, 3D Animation                                                         \\
P3          & UX Design, Product Design & 9                              & 1           & 1                          & 1              & Painting, Printmaking, Glitch Art,   Handicraft                   \\
P4          & Animation                          & 14                             & 3           & 4                          & 9              & Illustration, 2D Animation                                                              \\
P5          & Digital Media Arts                 & 10                             & 6           & 3                          & 8              & Motion Comics, Promotional Video,   
 Micro Movie, 3D Game, Theater     \\
P6          & Music Production                   & 13                             & 3           & 1                          & 2              & Dynamic Visuals and Sound-based Creation                                               \\
P7          & Transmedia Arts                    & 14                             & 4           & 1                          & 3              & 3D Game, Painting, Sculpture, Performance Arts, 2D/3D Animation          \\
P8          & Transmedia Arts                    & 8                              & 3           & 1                          & 1              & 3D Game, 3D Animation, Audio-Visual                                                        \\
P9          & Virtual Reality                    & 5                              & 3           & 1                          & 3              & Painting                                                                                \\
P10         & Transmedia Arts                    & 16                             & 6           & 1                          & 4              & Painting, 2D/3D Animation, Film                                                          \\
P11         & Advertising, Graphic Design        & 29                             & 4           & 1                          & 4              & Graphic Design, Sculpture, 2D Animation \\
P12         & Digital Media Technology           & 3                              & 2           & 1                          & 1              & Sculpture, Installation, Experimental Film, 3D Animation, 3D Game                     \\
P13         & Digital Media Technology           & 3                              & 1           & 1                          & 1              & Computer Graphics Art, 3D Game                                                                         \\
P14         & Virtual Reality                    & 6                              & 4           & 1                          & 1              & 3D Game                                                                                 \\
P15         & Motion Graphics                    & 7                              & 4           & 2                          & 2              & 2D Animated Shorts                                                                      \\
P16         & Graphic Design                     & 16                             & 3           & 3                          & 4              & Painting, Graphic Design, Video                                                         \\
P17         & Mural Arts                         & 20                             & 7           & 2                          & 8              & Kinetic Sculpture, Installation, Mural Arts, Computer Graphics Art                                     \\
P18         & Digital Media Arts, Animation      & 15                             & 4           & 2                          & 2              & Illustration                                                                            \\
P19         & Fine Arts                          & 10                             & 1           & 5                          & 10             & Illustration                                                                            \\
P20         & Digital Media Arts                 & 6                              & 3           & 2                          & 2              & Micro Movie, Photography, Painting, 2D/3D Animation, Installation                    \\
P21         & Motion Graphics                    & 24                             & 4           & 1                          & 1              & Illustration, GIF Animation, 2D Animated Shorts \\                               \bottomrule
\end{tabular}}
\end{table*}

\subsection{Authoring Tools for Animated VR Stories}
\label{sec:related_authoring_story}

Various authoring tools have been proposed in the literature for creating animated VR stories, which can be categorized into non-immersive, immersive, and hybrid based on the environments in which they are used. 

Non-immersive tools~\cite{henrikson2016storeoboard, zhao2020shadowplay2} can supplement existing commercial software (e.g., Blender and Unreal Engine), which is versatile but complex and oriented to broader 2D/3D game or animation creation. 
For example, Henrikson~\etal~\cite{henrikson2016storeoboard} designed an interactive tablet system for artists without 3D modeling skills to create stereoscopic storyboards with fluid pen-and-touch input.  
Similarly, ShadowPlay2.5D~\cite{zhao2020shadowplay2} allows novices to create 360-degree poetry stories, offering tailored features like an image repository and pen-based image animation. However, their 2D interfaces might cause a cognitive disconnect when considering spatial relationships and peripheral vision inherent in a 3D VR environment. 

These limitations drive interest in immersive tools, such as commercial tools (e.g., Quill and AnimVR) and research prototypes~\cite{galvane2019vr, stemasov2023sampling, wang2022videoposevr, nguyen2017collavr, vogel2018animationvr}. 
For example, Galvane~\etal~\cite{galvane2019vr} proposed a VR tool for storyboard creation that allows creators to arrange virtual spaces, capture snapshots, and then convert these into storyboards. This immersive approach gives creators a better understanding of spatial relationships.
AnimationVR~\cite{vogel2018animationvr} further allows direct manipulation via 6DoF controllers to animate characters, bypassing 2D gizmos.
Despite these advantages, immersive tools may bear accuracy issues, with limited feature support compared to their non-immersive counterparts.

To address these issues, research has explored hybrid tools~\cite{henrikson2016multi, lamberti2020immersive} to integrate the strengths of both classes. 
For example, VR Blender~\cite{lamberti2020immersive} combines the extensive support of Blender with the immersive authoring benefits of VR, thereby enhancing key animation tasks and facilitating reuse and modification. 
Henrikson~\etal~\cite{henrikson2016multi} proposed a multi-device system that facilitates artists to create storyboards on tablets while allowing directors to see them in VR.

While effective, these three types of tools remain fragmented and limited to certain tasks like storyboarding~\cite{henrikson2016multi, henrikson2016storeoboard, galvane2019vr}, asset animation~\cite{lamberti2020immersive, vogel2018animationvr, wang2022videoposevr}, and camera control~\cite{galvane2019vr}. 
Creators need to perform many other tasks~\cite{gipson2018disneycicles} to create animated VR stories and desire tools that blend into their creation processes~\cite{cutler2019making}.
To inform future research in creativity support for VR stories, we provide a nuanced understanding of the creation processes and challenges based on interviews with creators using both non-immersive and immersive tools.

\section{Interview Study Design and Analysis}
To answer RQ1 and RQ2, we conducted semi-structured interviews with 21 creators to collect qualitative and in-depth insights on their creation processes and challenges. 

\subsection{Positionality Statement}
The study design, data collection, and data analysis were mainly performed by five authors with interdisciplinary backgrounds (denoted as A1--A5). A1 has 4-year experience in HCI research and 3-year experience in developing VR with game engines. A2 has 13 years of experience in digital entertainment development and production, sophisticated in 3D animation, VR shorts, and interactive films. A3 is an award-winning animation artist with 11 years of experience in directing 2D/3D animation and 3 years of experience in VR storytelling practice and research. A4 and A5 are technical HCI researchers with 6 and 15 years of experience, respectively.

\subsection{Recruitment and Interviewees}
\label{sec:interview_recruitment}
We recruited creators in various ways and interviewed those who had crafted at least one animated VR story.
First, we searched for VR stories on popular local content platforms and found the corresponding creators. We sent eleven interview requests and obtained five acceptances. 
Second, we distributed our advertisement to several art schools and received seven qualified responses.
Lastly, we reached out to nine creators from our personal network.

We stopped recruiting once interviewees' insights converged, ending up with 21 creators (all of them Chinese; 13 females and 8 males; age groups: 22--27 (14), 28--33 (4), 34--39 (2), 40--46 (1)). Table~\ref{tab:participant_information} lists their professional backgrounds and experiences. Specifically, all interviewees had formal art training but came from different subfields, such as digital media arts and graphic design. Their experience in non-VR art and VR art averaged 11.8 (min=3, max=29) years and 3.4 (min=1, max=7) years. On average, they created 1.8 (min=1, max=5) VR stories, spending approximately 3.4 (min=0.5, max=12) months per story. 
They also had abundant experience in various types of VR artworks (e.g., VR paintings) and non-VR artworks (e.g., 2D/3D animation, desktop games, and films).

\RR{In terms of their sociocultural backgrounds, fourteen interviewees were studying or working in various cities across China, while the others were in the United States, the United Kingdom, Japan, and Finland. Their creative works reflected a blend of cultural influences, incorporating both traditional and contemporary elements. Notably, Chinese traditional culture (N=16) serves as a foundational influence, encompassing classical literature (e.g., Tang poetry), religious philosophies (e.g., Buddhism and Taoism), and ethnic minority art. There is also a strong influence from anime culture (N=5), popular culture (N=5), science fiction (N=3), and gaming culture (N=2).}

\subsection{\RR{Question Design} and Data Collection}

\RR{We developed our interview questions through brainstorming, refinement, and testing. Starting from our two research questions, A1 and A2 independently brainstormed potential interview questions and then collaborated to organize them.
A1 then refined the questions with A4 and A5. To test the questions, A1 conducted a mock interview with A3, an animated VR creator. A3's feedback highlighted the need to contextualize questions with concrete projects to elicit insights, which led us to incorporate project-specific questions. To control interview duration, we divided the resulting questions into two parts: (1) a pre-interview questionnaire (Appendix~\ref{appendix:questionnaire}) that included project-specific questions about interviewees' favorite VR stories, and (2) an interview guide (Appendix~\ref{appendix:guide}) to facilitate the discussion of creation processes and associated challenges.
}

The data collection with each interviewee consisted of three steps. First, we obtained their consent to join the interview and be video and audio recorded. Thirteen interviewees permitted us to use their intermediate and final artwork in our manuscript for research purposes, with their preferred pseudonyms included in the figure captions. Next, they completed the questionnaire at least two days before the interviews and prepared demos and materials related to their crafted stories. Detailed information about their favorite VR story projects can be found in Appendix~\ref{appendix:project}. Finally, we conducted individual semi-structured interviews \RR{online via video conferencing software}. The interview sessions averaged two hours, and all audio recordings were automatically transcribed for analysis.

\subsection{Interview Data Analysis}
In terms of \textbf{RQ1} about creation processes, we analyzed the interview data in three steps. 
Firstly, A1 extracted stages from each interviewee's descriptions, created a flow map to show their sequence, and annotated each stage with its inputs, outputs, and supporting quotes on Miro whiteboards. This process resulted in 21 whiteboards, each corresponding to one animated VR story. 
Secondly, A2 and A3 went through each whiteboard and discussed together with A1. For each whiteboard, we examined whether any mistakes existed in A1's initial organization and checked whether some original stages mentioned by the interviewees could be merged or split. 
\RR{For example, we merged technical stages like modeling, texturing, and rigging into a broader stage of asset development, as they collectively create usable 3D assets.}
Thirdly, we unified stage names across all whiteboards. 
\RR{Since our interview protocol did not prescribe any specific stages, the interviewees often used different terms to refer to the same stage. For example, \qq{whiteboxing} and \qq{3D layout} were both used to describe the previsualization practices. To determine appropriate stage names, we first referenced 3D animation~\cite{beane20123dAnimation} and filmmaking~\cite{jeffrey2020ves}, adopting terms such as \qq{idea}, \qq{story}, \qq{script}, \qq{storyboard}, \qq{previsualization}, and \qq{design}. We then refined these terms to better align with our collected data. For instance, we specified the design stage as \qq{visual, audio, and interaction design}. Additionally, we introduced the terms \qq{scene assembly} and \qq{story integration} to describe interviewees spatially and temporally composited all story elements.}
A1, A2, and A3 iterated between the second and third steps until we were satisfied with the results and then discussed them with A4 and A5.

To address \textbf{RQ2} about challenges, we conducted a thematic analysis~\cite{braun2019reflecting}. A1, A2, and A4 independently went through \RR{the same transcripts from four interviewees (approximately 20\% of all interview data}), highlighting sentences related to challenges and generating initial codes. Utilizing affinity diagrams, A1, A2, and A4 compared, discussed, and grouped their codes into themes. The themes were then reviewed together with A5 to develop an initial codebook. Subsequently, A1 coded the remaining transcripts using this codebook \RR{while maintaining sensitivity to new themes. When potential new themes emerged, A1 documented relevant instances and discussed them with A2 and A4 to refine the codebook.}
The final coding results were reviewed and refined through group discussions with all researchers.

\section{Creation Process}
\label{sec:processes}

This section reports findings on the creation processes of animated VR stories (\textbf{RQ1}), including individual stages and overall workflows.

\begin{figure*}[!tb]
    \centering
    \includegraphics[width=\linewidth]{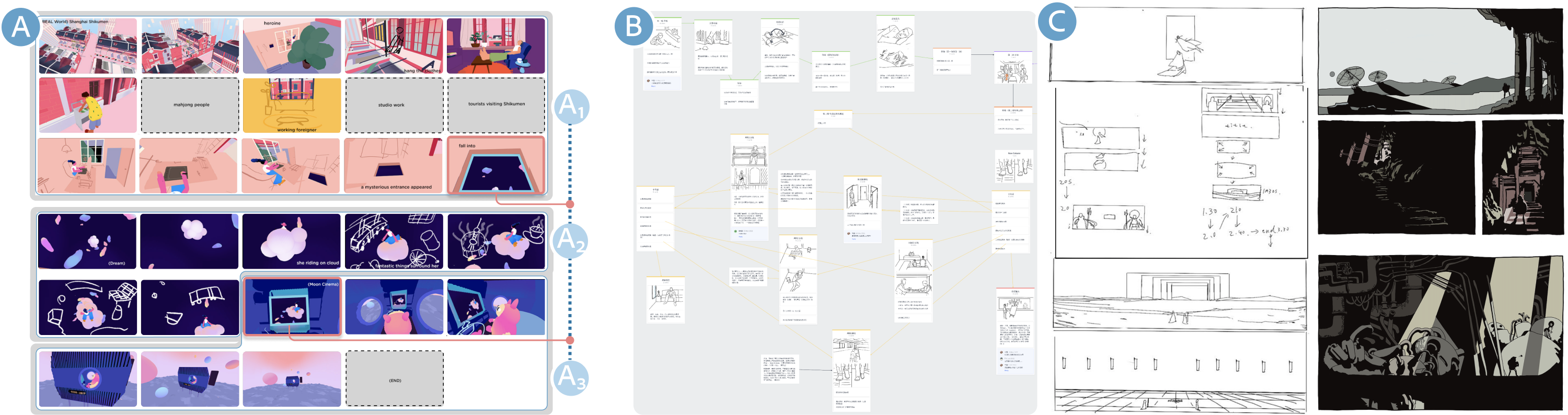}
    \caption{Various practices and purposes in storyboarding. (A) A high-fidelity storyboard for post-hoc refinement (\CR{\copyright AMAO}). The creator inspected the compact overview of three world settings (A1-A3) and directly inserted placeholders or added sketches on areas to enhance. (B) A storyboard with interconnected nodes detailing the entire storyline for team communication (\CR{\copyright Hedi's team}). (C) A storyboard in two forms by a solo creator that selectively captures some moments for self-evaluation (\CR{\copyright Ocean Hu}).} 
    \label{fig:storyboarding}
    \Description{This figure showcases different practices in storyboarding. Part A presents a colorful storyboard sequence progressing from everyday life to dreamlike and futuristic scenarios. Part B illustrates a mind-map style diagram connecting various narrative elements and scenes with sketches. Part C contains rough black-and-white storyboard panels, showing a character's journey through outdoor landscapes using minimal color and shading for depth.}
\end{figure*}

\begin{figure*}[!tb]
    \centering
    \includegraphics[width=\linewidth]{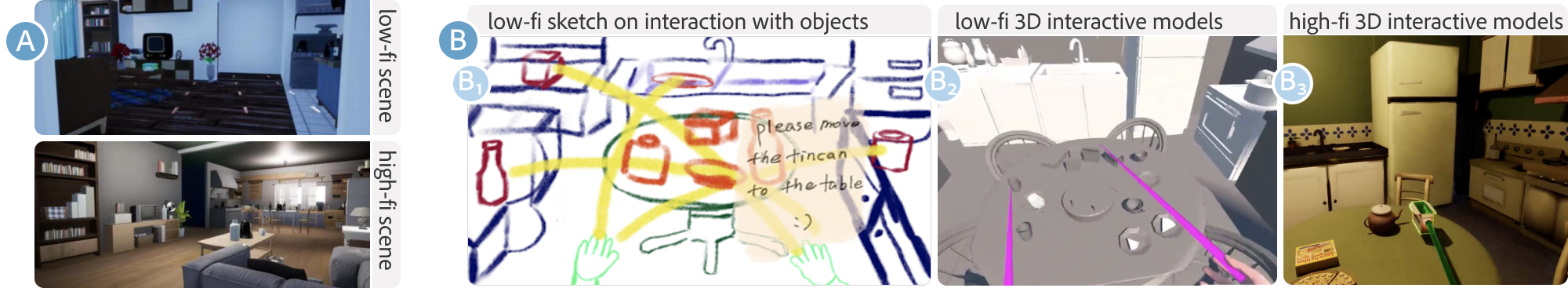}
    \caption{Examples of previs to assess scenes and the functionality of interactions before developing high-fidelity models. (A) Low-fidelity models in the previs stage and high-fidelity scenes in the ultimate story (\CR{\copyright Hedi's team}). (B) Interactive elements that (B1) started as sketches, (B2) evolved into low-fidelity prototypes, and (B3) were incorporated into the final high-fidelity story (\CR{B1-B3 \copyright Coin's team}).} 
    \label{fig:previsualization}
    \Description{This figure showcases examples from the previsualization stage. Part A compares low-fidelity and high-fidelity scenes. Part B focuses on object interaction, starting with a low-fi sketch (B1), followed by low-fi 3D models (B2), and concluding with high-fi 3D interactive models (B3), illustrating the evolution from rough concepts to fully interactive, realistic environments.}
\end{figure*}

\subsection{Common Stages}
\label{sec:common_stages}

Our interviewees went through 10 common stages to transform their ideas into an animated VR story. 
Figure~\ref{fig:workflow_example_1} illustrates a concrete example. 
For brevity, we use the pre-production phase and production phase when discussing multiple stages later. Specifically, the pre-production phase refers to the six stages (Sec.~\ref{sec:idea_generation}-\ref{sec:design}) from idea generation to design, where an animated VR story is conceptualized and planned. The production phase refers to asset development (Sec.~\ref{sec:asset_development}), scene assembly (Sec.~\ref{sec:scene_assembly}), and story integration (Sec.~\ref{sec:story_integration}), marking the hands-on execution that brings the story to life. Our interviewees did not have post-production activities like editing before conducting the evaluation (Sec.~\ref{sec:evaluation}).
Below describe each stage and summarize corresponding practices.

\wrap{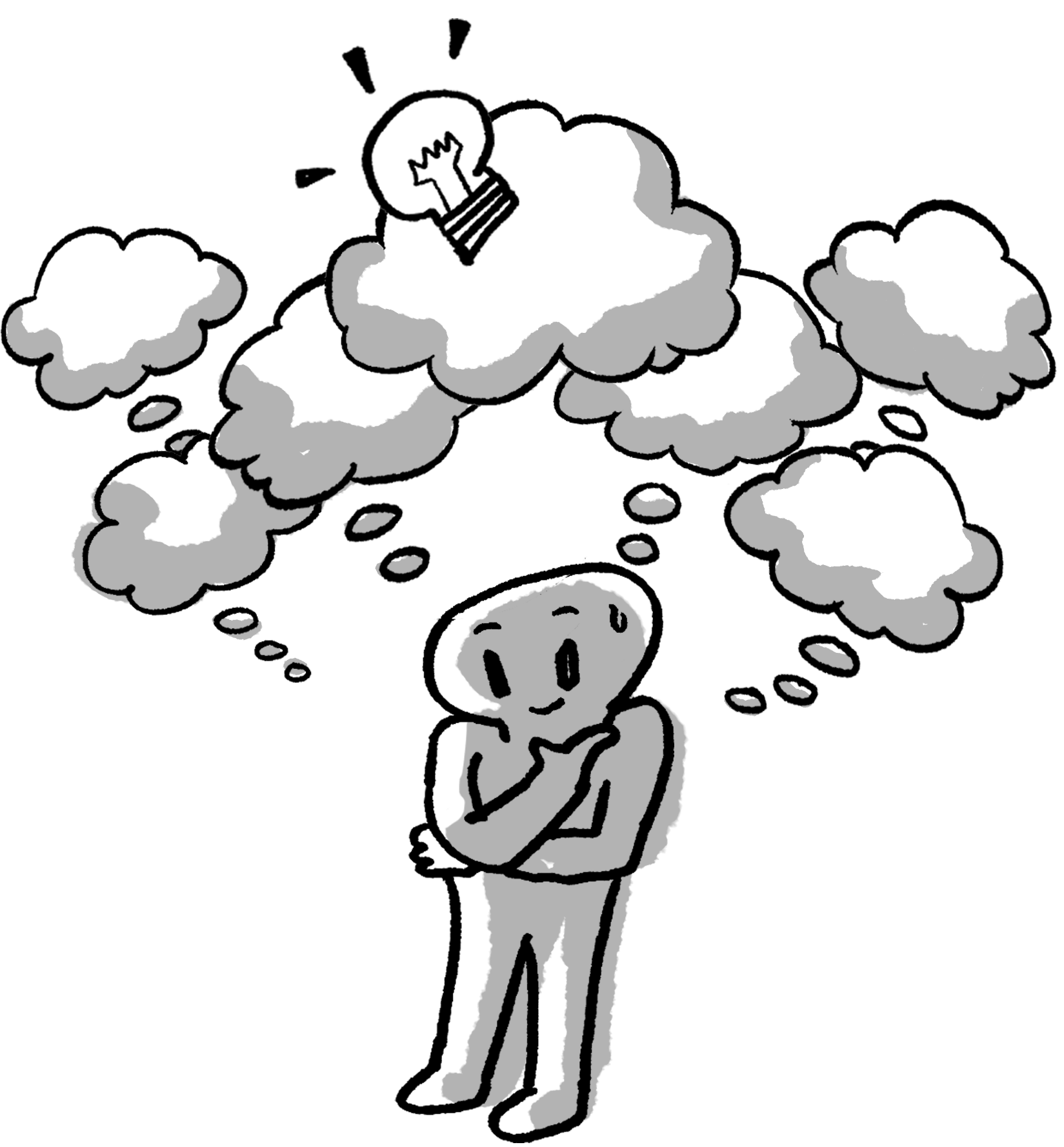}
\subsubsection{Idea Generation}
\label{sec:idea_generation}
Creators draw their initial ideas from various sparks, such as personal experiences, dreams, and existing artworks. They then think about what themes, messages, and emotions to convey. 
They may select visual styles (e.g., realistic versus cartoonish), set an overall mood or atmosphere, and have rough pictures of characters. 
Creators may assess if VR is the right medium for their vision, making decisions on aspects that require particular attention in this medium, such as 3-DoF or 6-DoF camera control, passive or active viewer roles, levels of interactivity, and the incorporation of multiple perspectives. 

\wrap{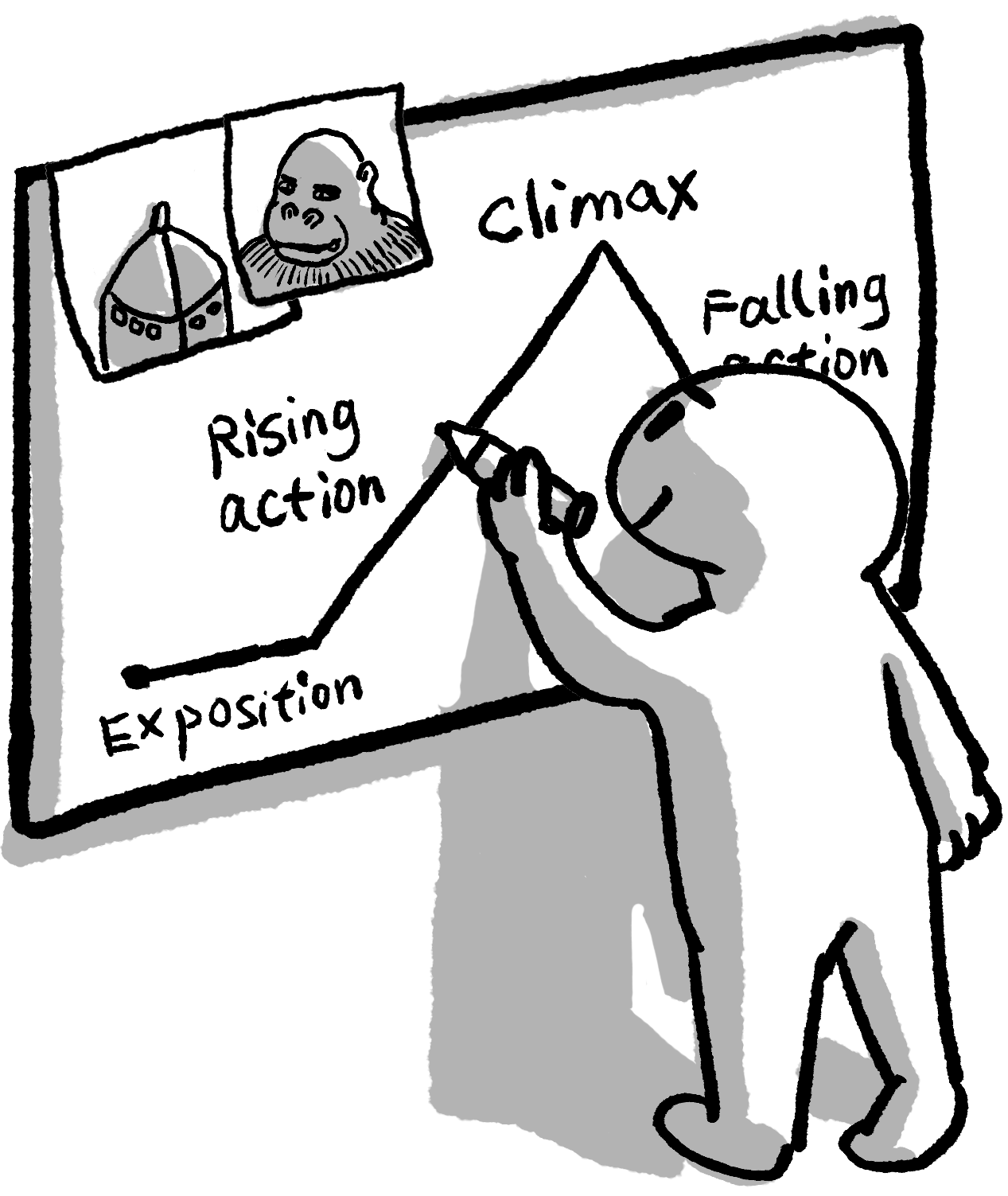}
\subsubsection{Story Creation}
\label{sec:story_creation}
Creators outline the backbone of a VR story and determine what viewers would experience sequentially (see Fig.~\ref{fig:workflow_example_1}-B1). They also create the main story elements, such as characters, worlds, and storylines (see Fig.~\ref{fig:workflow_example_1}-B2). Our interviewees employed direct (e.g., linear narrative) and/or indirect (e.g., leaving clues in VR environments) ways to tell their stories.

When aiming for maximal clarity in conveying messages to the viewers, interviewees adopted a more direct approach. They prioritized the development of characters and central conflicts. Most interviewees adopted a linear narrative structure and organized the story's events in chronological order. Only P16 incorporated a branching narrative, offering viewers the opportunity to influence the story's direction through choices of their roles in the story.

Conversely, when aiming for an open-ended and exploratory story experience, interviewees preferred indirect approaches, mainly using environmental storytelling to exploit VR's immersive and multi-sensory benefits. As an example, P2 shared, \q{Inspired by a science fiction novel, I used five interconnected environments to indirectly tell my story. Each environment left subtle clues, such as sun positioning for a timeline and color changes for different places. I allowed viewers to explore freely to feel and find these clues and form their own interpretation of the story.}

\wrap{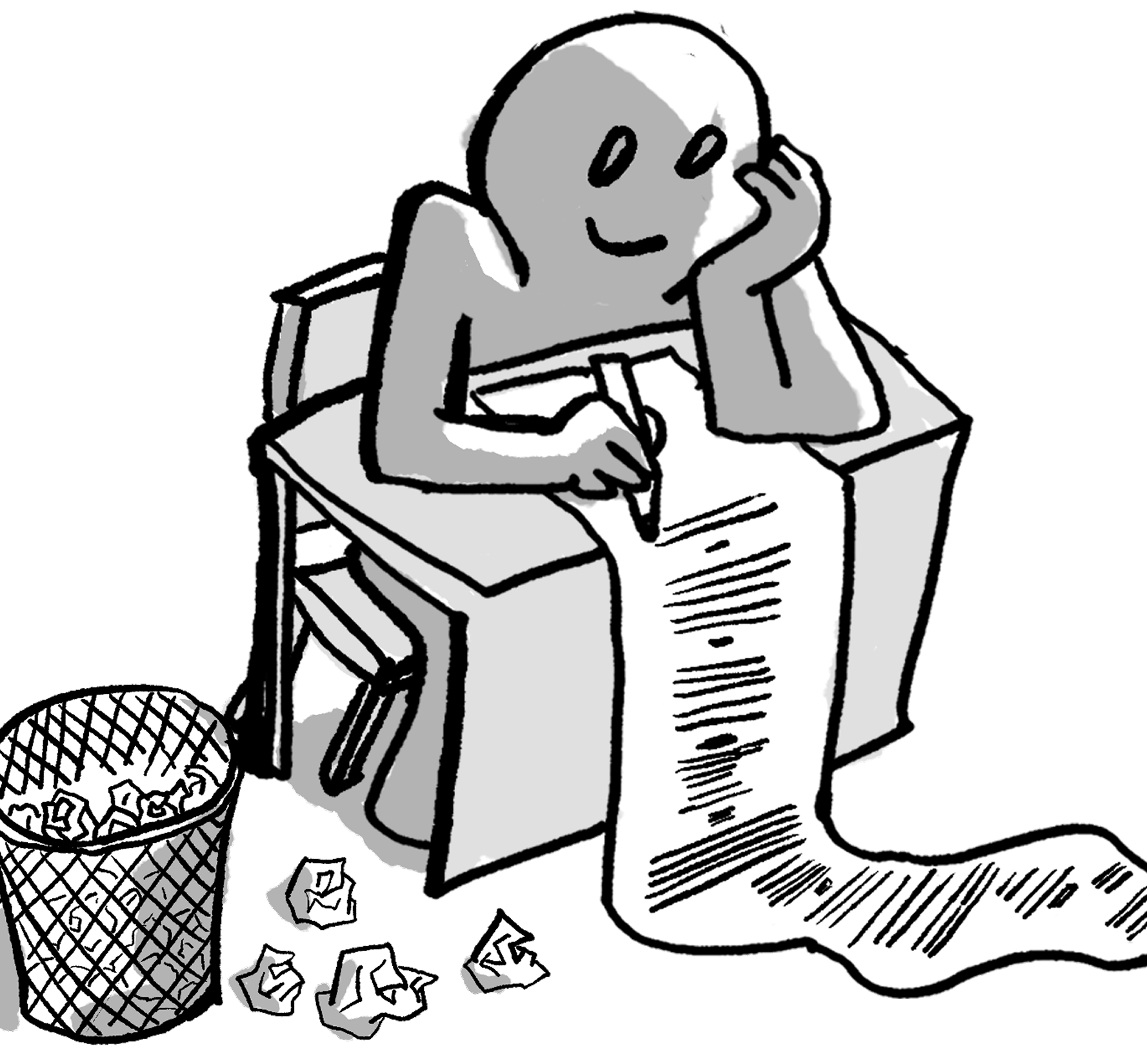}
\subsubsection{Scriptwriting}
\label{sec:script_writing}
Scripts are a written and structured format of stories. Interviewees' scripts varied. Some might only include common elements like character dialogue and narration, and others might also include audiovisual cues (e.g., camera setting, sound effects) and interactions at different levels of detail. 

\wrap{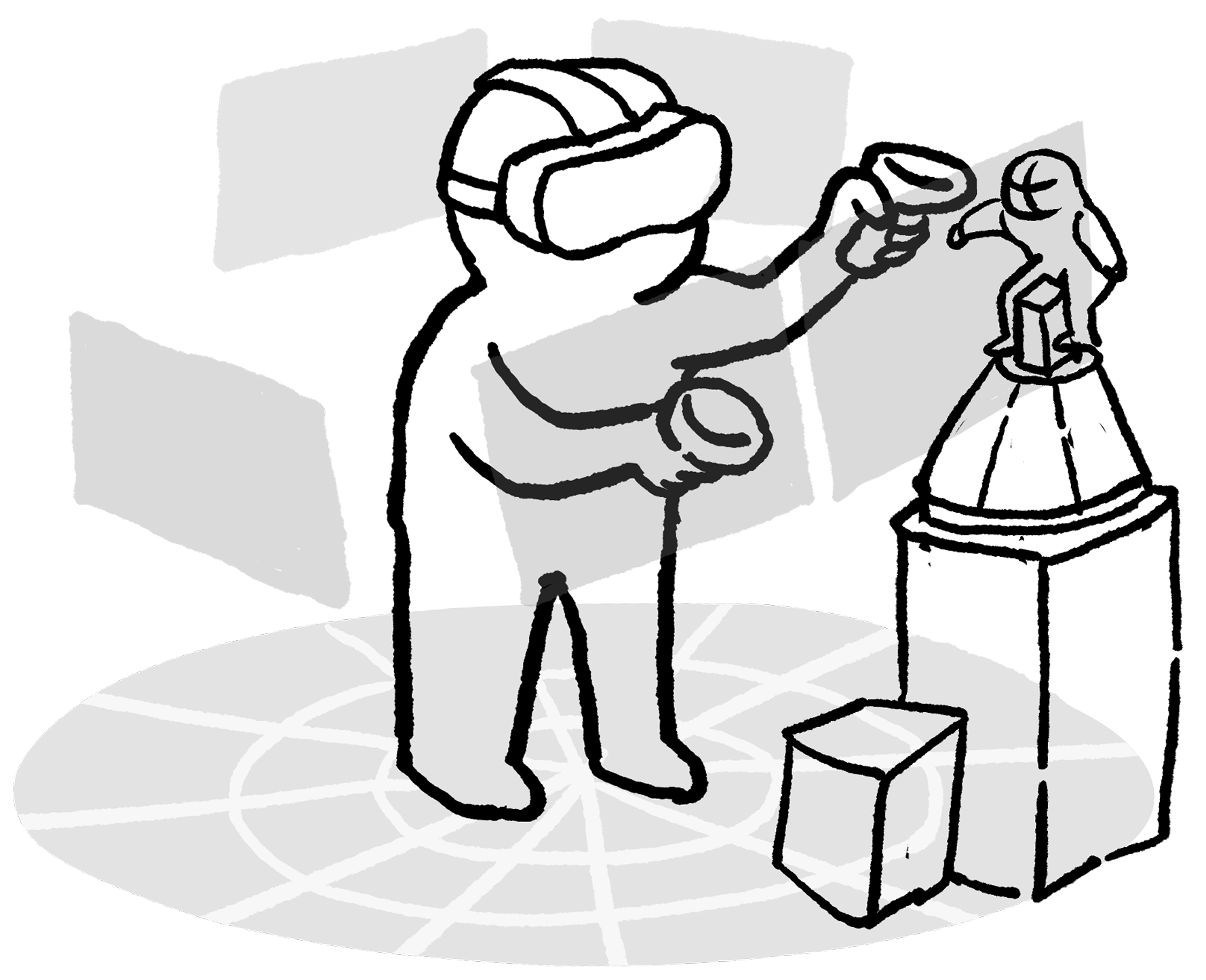}
\subsubsection{Storyboarding}
\label{sec:storyboarding}

\begin{figure*}[!tb]
    \centering
    \includegraphics[width=0.98\linewidth]{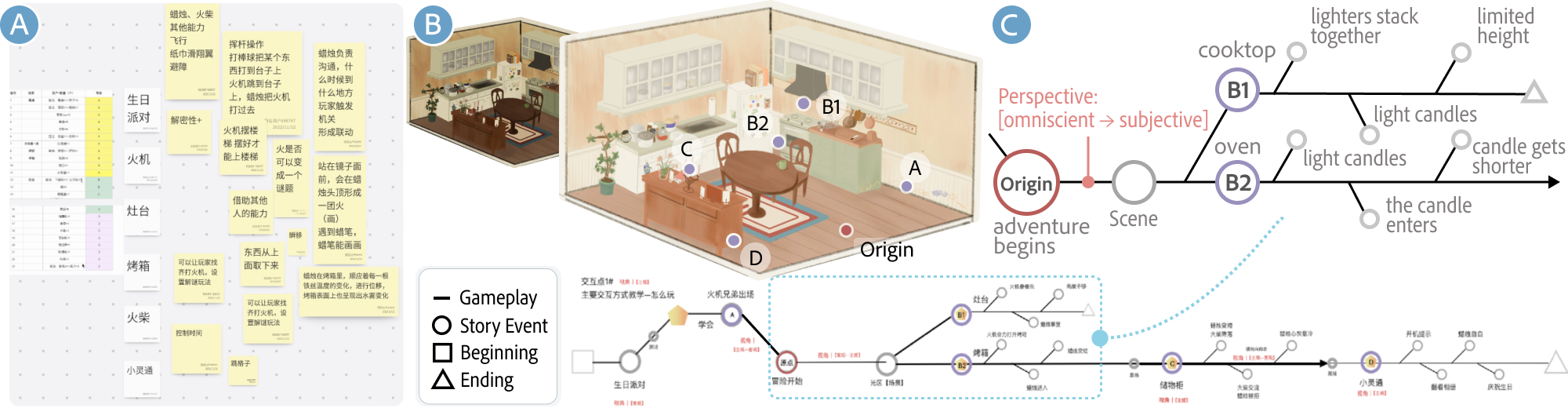}
    % \vspace{-8mm}
    \caption{An example of interaction design while connecting both the story plots and scenes (\CR{\copyright Coin's team}). (A) Brainstorming on a whiteboard about possible interactions with available objects within a scene. (B) Marking places where interactions will be triggered on a scene image. (C) Representing interactions in a fishbone diagram to examine whether these interactions can propel the plots forward.}
    \label{fig:interaction_design_1}
    \Description{This figure demonstrates how interactions are designed in relation to story plots and scenes. Part A shows a structured plan with story events and gameplay elements arranged in columns and sticky notes. Part B presents a 3D kitchen scene, highlighting interaction points such as the oven (B2) and cooktop (B1), where specific actions occur. Part C illustrates a flowchart of story progression, moving from an omniscient to a subjective perspective as the adventure begins, with interactions like lighting candles and managing objects integrated into the plot.}
\end{figure*}

\begin{figure*}[!tb]
    \centering
    \includegraphics[width=\linewidth]{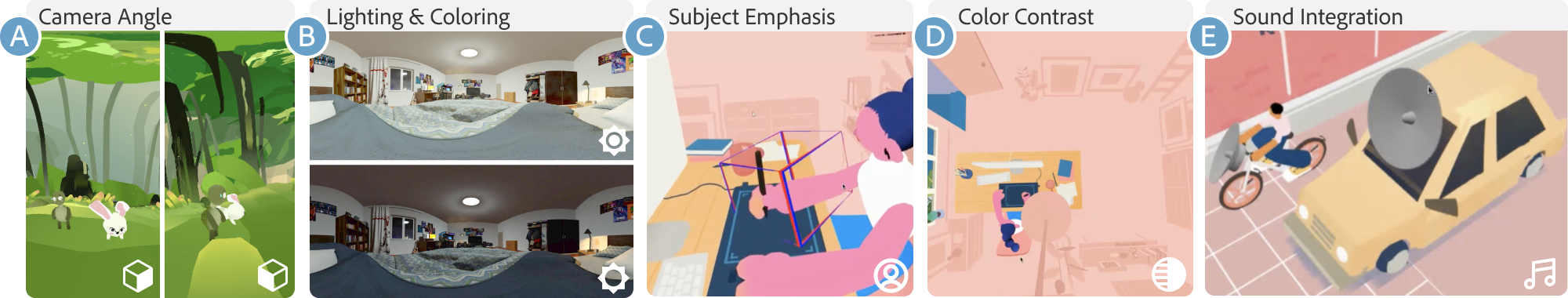}
    % \vspace{-8mm}
    \caption{Key considerations in the scene assembly stage. (A) Optimizing the coherence of 3D spatial relationships amongst assets across varied camera angles (\CR{\copyright Jiaming's team}). (B) Calibrating light and color settings to convey subtle ambiance (\CR{\copyright Tiemu}). (C-E): Guiding viewers' attention by (C) enlarging the subject of interest, (D) creating color contrast, and (E) integrating sound effects (\CR{C-E \copyright AMAO}).} 
    \label{fig:scene_assembly}
    \Description{This figure highlights key considerations in the scene assembly stage. Part A shows a scene with a rabbit and a turtle from different camera angles. Part B compares two versions of the same scene to show how lighting and coloring affect mood. Part C emphasizes the subject, demonstrating how framing draws attention to key actions. Part D explores color contrast. Part E integrates sound, linking it to visual elements, such as a vehicle with a satellite dish.}
\end{figure*}

Storyboards are a series of sketches that visualize the story flow, key moments, and plots.  Our interviewees configured initial scene compositions, camera settings, and character poses in storyboards. Because viewers control cameras in VR, interviewees also needed to guess how viewers would observe and interact with the scenes and then annotated visual and auditory cues to direct viewers' attention. As shown in Fig.~\ref{fig:storyboarding}, our interviewees used storyboards in different ways for various purposes.

\wrap{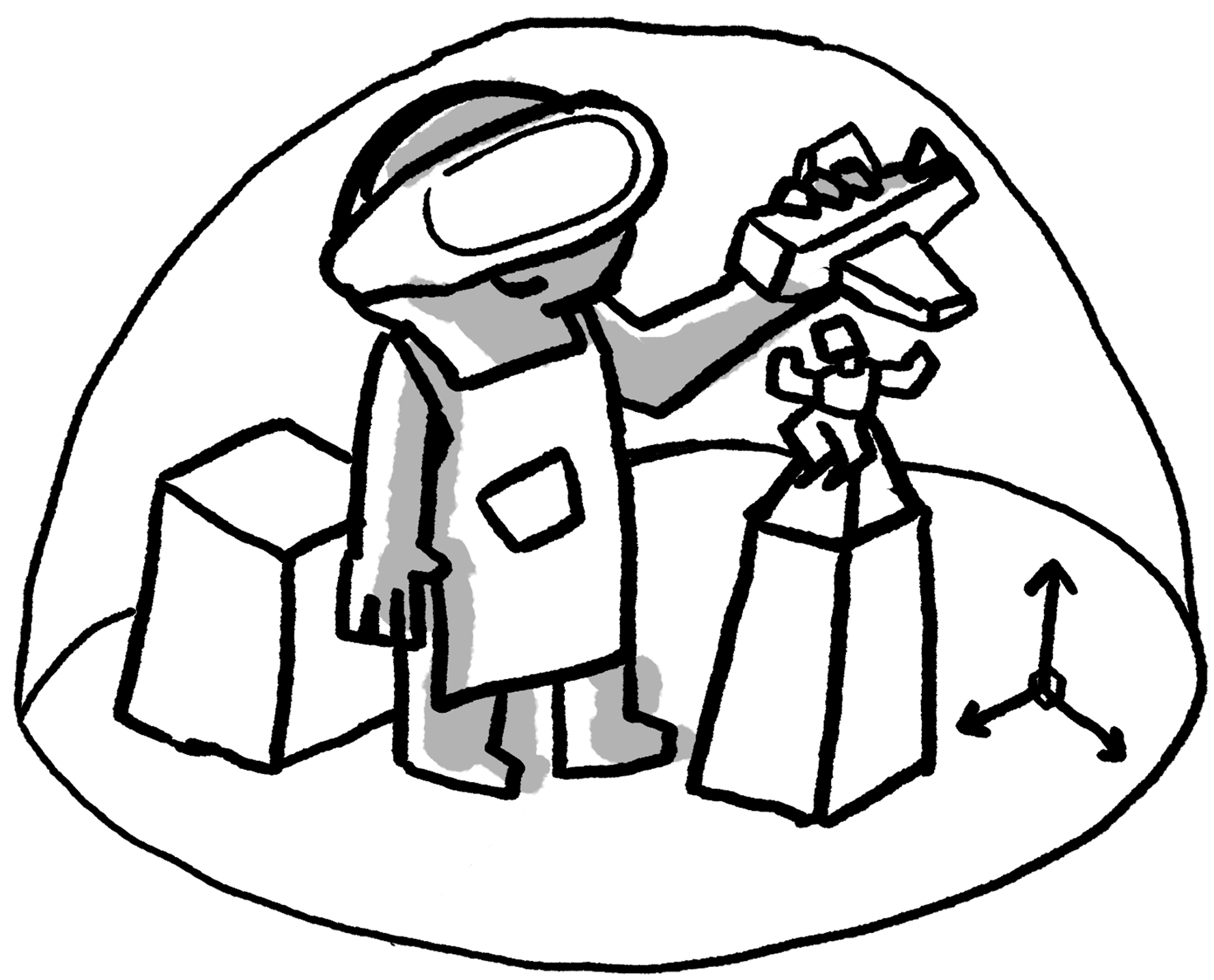}
\subsubsection{Previs}
\label{sec:previs}
Previs (or previsualization), involves the use of rough 3D visual representations for story segments or whole stories, previously described in natural language or 2D pictures. 
As shown in Fig.~\ref{fig:previsualization}, 
low-detail 3D models were laid out in 3D spaces to represent scenes and interactions, serving as proxies for the final, polished versions. 
A few interviewees used previs selectively for key story segments. They focused on exploring spatial relationships, experimenting with dynamic factors like camera and character movement, and prototyping VR interactivity. 
Some interviewees from larger teams used previs to map out the entire stories to aid the team and stakeholder communication.
For example, P4 acted as a director and utilized Quill to create comprehensive previs that conveyed her vision to Unity developers without art backgrounds and helped secure approval from her superiors.

\wrap{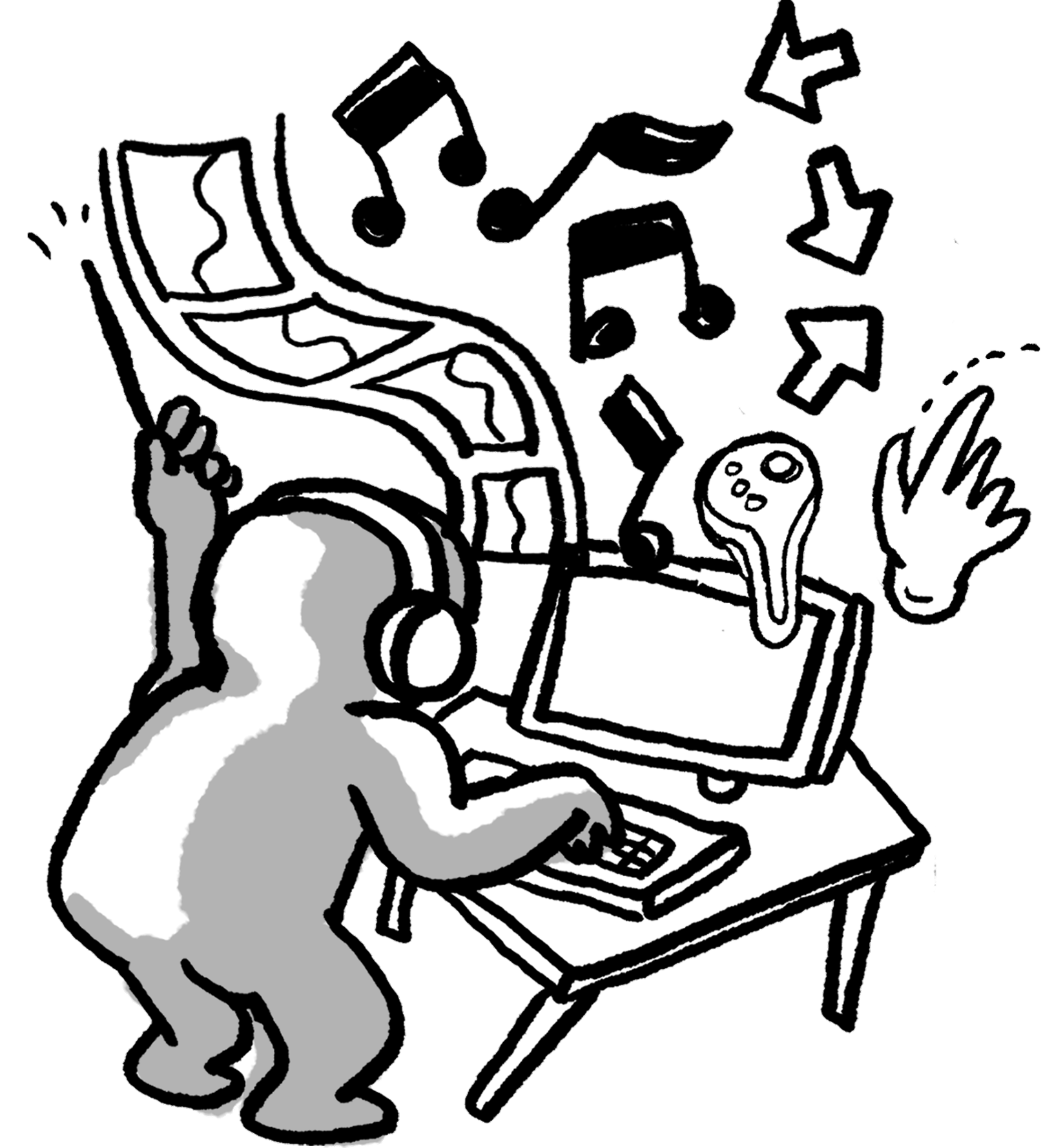}
\subsubsection{Visual, Audio, and Interaction Design}
\label{sec:design}

Creators transform their imagination and visions into physical or digital representations (e.g., drawings and videos) of visual, auditory, and interactive elements. 
Our interviewees designed with VR's characteristics in mind, building upon previous settings on the story's themes, mood, and atmosphere.
For visual design, they used drawings to establish the look of the characters, props, and environments. 
For audio design, they determined the rhythm and functions of background music, sound effects, and spatial audio that reinforced the story's emotional beats and atmosphere.
For interaction design, they particularly cared about whether the interaction enhanced viewers' engagement rather than distracting viewers from the story. P9 stated, \q{If an interaction was isolated or distracted viewers, I would leave it out.} Thus, our interviewees thought about where and when to introduce interactions based on scenes and storylines in order to connect viewers with the scene and propel the plot forward. Figure~\ref{fig:interaction_design_1} shows an example in which a group of creators brainstormed and annotated where interaction happened, and used a fishbone diagram to integrate interaction and plots.

\wrap{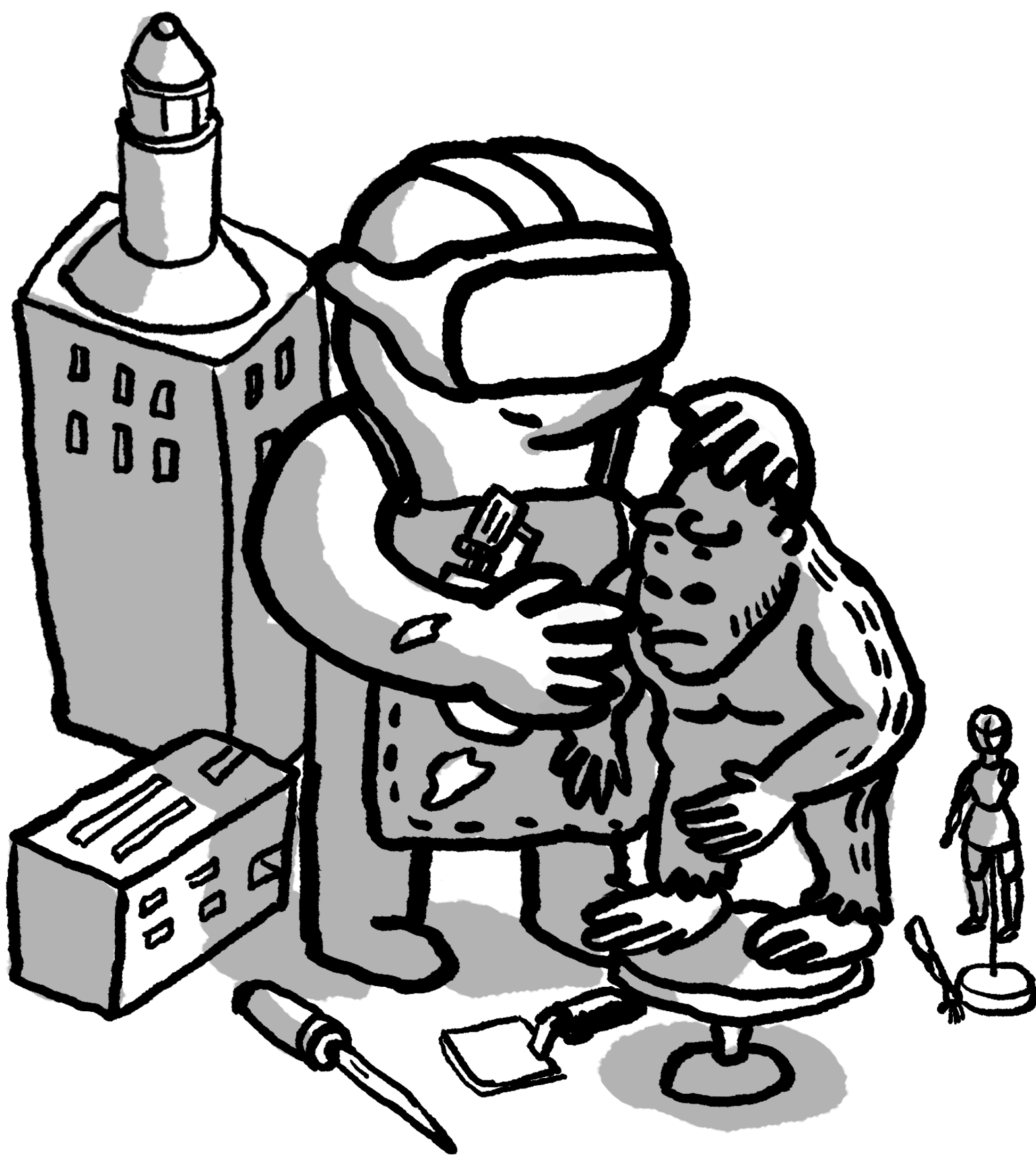}
\subsubsection{Asset Development}
\label{sec:asset_development}

Creators develop 3D models, textures, rigs, and animations for characters, props, and environments, transforming the design concepts into digital assets. They also produce auditory elements like music and voice-over and implement interaction logic for interactive elements.
Our interviewees dedicated much effort to achieving specific visual aesthetics, either as an overall style (e.g., \q{ink painting} by P1) or for individual elements (e.g., \q{sparkling water} by P2 and \q{mysterious mist} by P7).

\begin{figure*}[!tb]
    \centering
    \includegraphics[width=\linewidth]{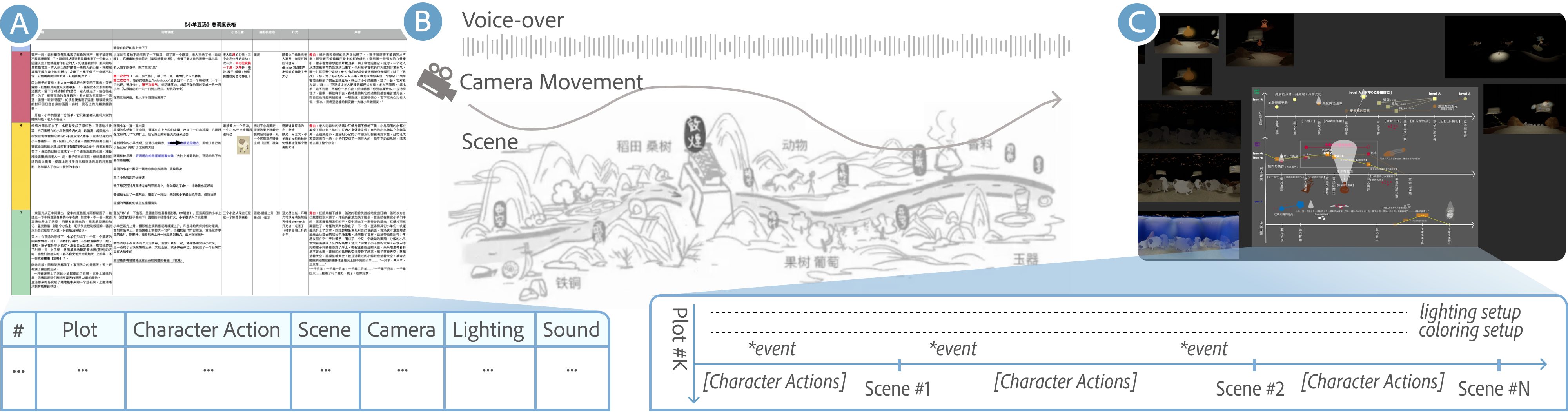}
    \caption{Practices of within-scene integration that synchronizes multiple story elements temporally. (A) Listing by-plot set-ups with a table, which covers character actions, scenes, and configurations of the camera, light, and sound (\CR{\copyright Coin's team}). (B) Aligning the camera and voice-over with the scene to guide viewers on a tour through various cities (\CR{\copyright ZhangXS}).  (C) Illustrating how lighting should change along a timeline with marked events (\CR{\copyright Coin's team}).} 
    \label{fig:within_scene_integration}
    \Description{This figure illustrates practices for within-scene integration that coordinate multiple story elements over time. Part A presents a detailed chart mapping out the plot, character actions, scenes, camera angles, lighting, and sound. Part B demonstrates camera movement and voice-over integration, with a scene flow depicted visually to show the progression of events. Part C highlights the technical setup for lighting and coloring, displaying a timeline that links character actions and scene transitions.}
\end{figure*}

\begin{figure*}[!tb]
    \centering
    \includegraphics[width=\linewidth]{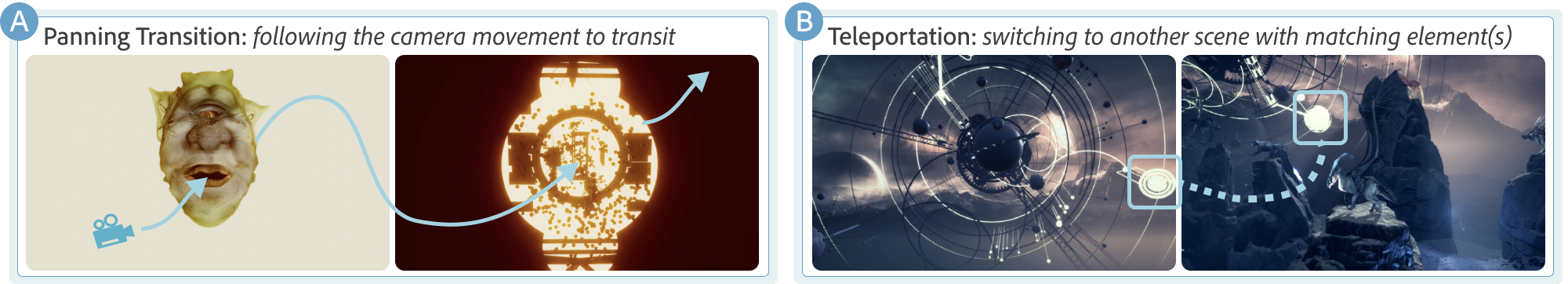}
    \caption{Practices of between-scene integration that concern how a scene transits to another (\CR{\copyright Angela Cai}). (A) Panning transition: cameras brought viewers to mythical worlds by moving through a mythical monster's mouth and a keyhole. (B) Teleportation: a rotating turntable teleported viewers between multiple parallel universes through similar visual elements.} 
    \label{fig:scene_transition}
    \Description{This figure illustrates practices for between-scene integration, focusing on how one scene transitions to another. Part A showcases a "Panning Transition," where camera movement smoothly follows a subject to transition between scenes. Part B demonstrates "Teleportation," where the transition occurs by switching to another scene with matching elements.}
\end{figure*}

\wrap{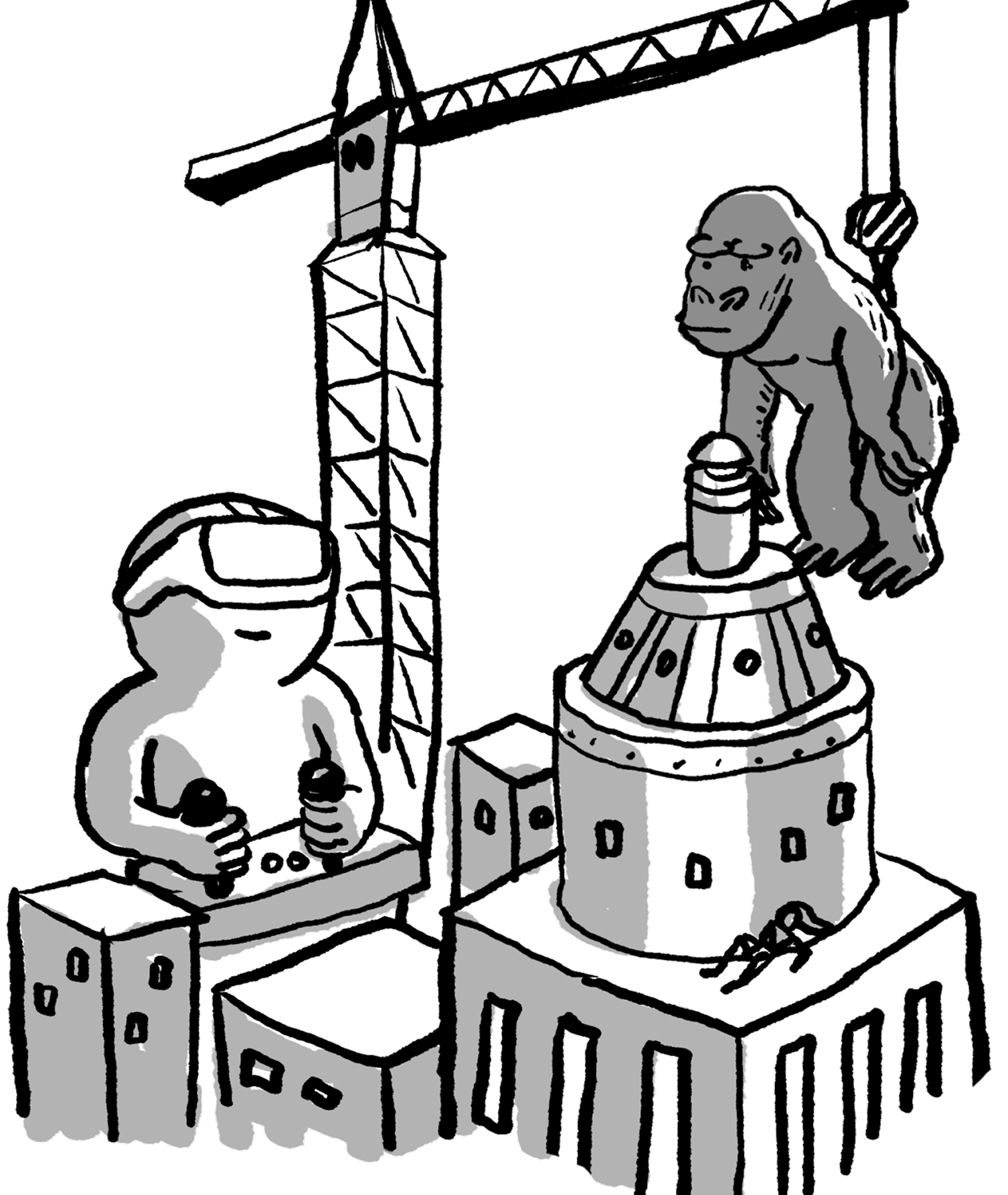}
\subsubsection{Scene Assembly}
\label{sec:scene_assembly}

The scene assembly stage focuses on the spatial aspects of the story. 
With individual assets (e.g., characters and props) at hand, our interviewees assembled these elements in 3D spaces to construct narrative scenes. Relying on prior storyboards and previs, they undertook multiple iterations to fine-tune the placement of assets, cameras, lights, and sounds. The placement involved several key considerations (Fig.~\ref{fig:scene_assembly}) for both aesthetic cohesion and narrative efficacy. 
First, they carefully considered the 3D spatial layouts amongst assets from multiple angles (Fig.~\ref{fig:scene_assembly}-A), such as placing assets at different layers to create depth. 
Second, they fine-tuned lighting, color, and special effects to craft the desired atmosphere and aesthetic within individual scenes. For instance, an interviewee adjusted a room scene to be darker and dimmer to express a feeling of depression (Fig.~\ref{fig:scene_assembly}-B). Third, they paid particular attention to guiding viewers' attention by experimenting with various audiovisual cues. For example, they enlarged the points of interest as an alternative to close-up camera shots (Fig.~\ref{fig:scene_assembly}-C). They might also simplify surroundings but emphasize the points of interest with colors (Fig.~\ref{fig:scene_assembly}-D) or sounds (Fig.~\ref{fig:scene_assembly}-E).

\wrap[1.45cm]{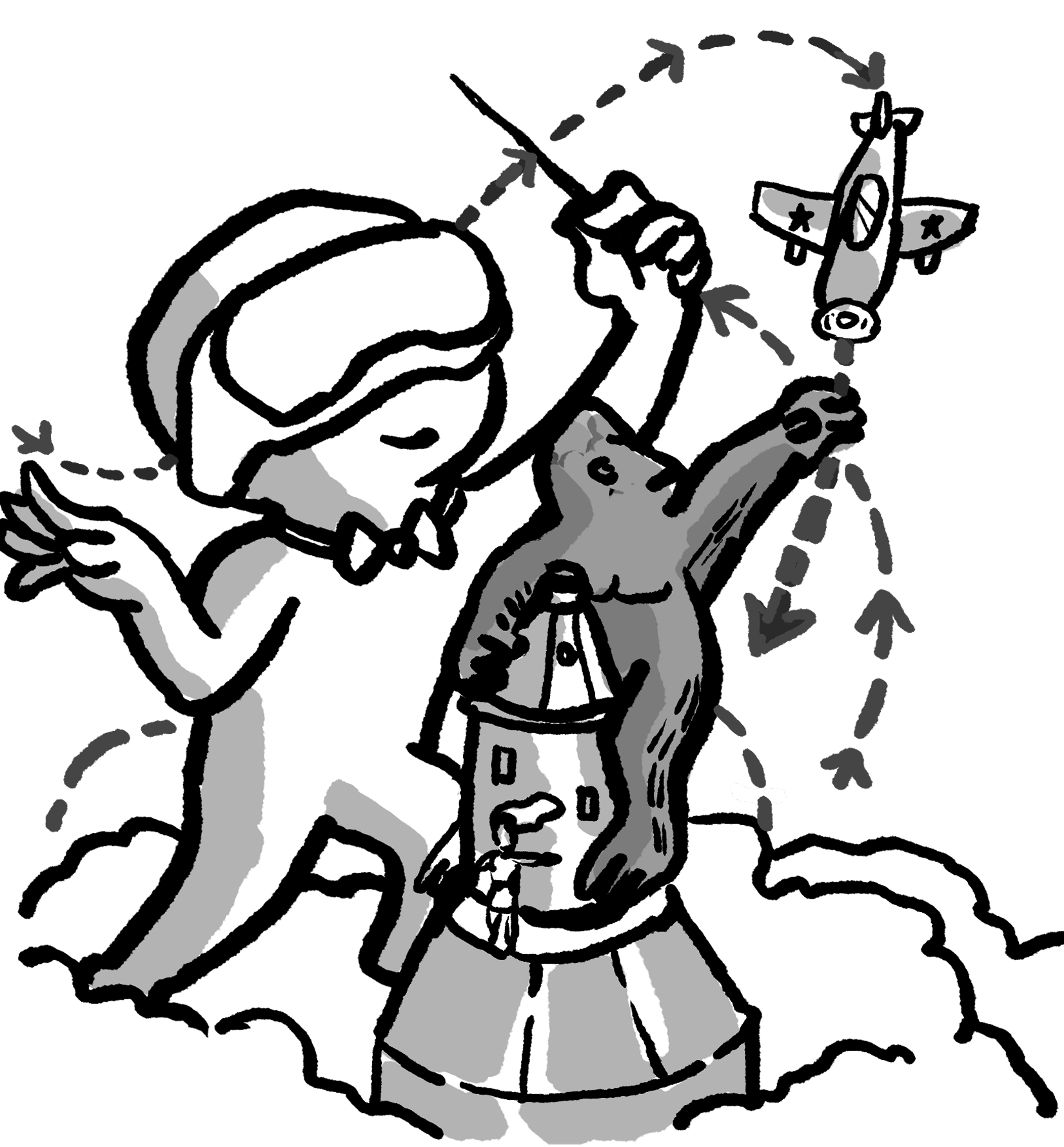}
\subsubsection{Story Integration}
\label{sec:story_integration}

With the spatial stages already set, this stage primarily involves the temporal orchestration (e.g., timing, duration, and sequencing) of story elements within and between scenes to form a complete story. This stage consists of within-scene integration (Fig.~\ref{fig:within_scene_integration}) and between-scene integration (Fig.~\ref{fig:scene_transition}).

For within-scene integration, interviewees engaged in two key tasks. First, they gave temporal dynamics to story elements such as characters, lighting, and cameras by specifying their movements, which included trajectory, duration, and speed. Second, they coordinated and synchronized these dynamic elements to construct plots. They strove to make all the elements flow together cohesively over time to deliver the intended messages or emotions. For example, an interviewee used a table (Fig.~\ref{fig:within_scene_integration}-A) to detail how each element should move in each plot with timestamp and duration specifications. The interviewee also used an illustration (Fig.~\ref{fig:within_scene_integration}-C) to think about how lighting should change along events. 
Another interviewee aligned the camera and voice-over with the environment (Fig.~\ref{fig:within_scene_integration}-B) to guide viewers on a tour through various cities.

During between-scene integration, interviewees worked on connecting individual scenes smoothly. They often chose transition methods that seamlessly fit into the story's context and utilized unique experiences provided by VR. For example, an interviewee adopted two transition methods: one used panning cameras to transition from one scene to another (Fig.~\ref{fig:scene_transition}-A), while the other offered teleportation experiences with matching elements (Fig.~\ref{fig:scene_transition}-B).

\begin{figure*}[!tb]
    \centering
     \includegraphics[width=\linewidth]{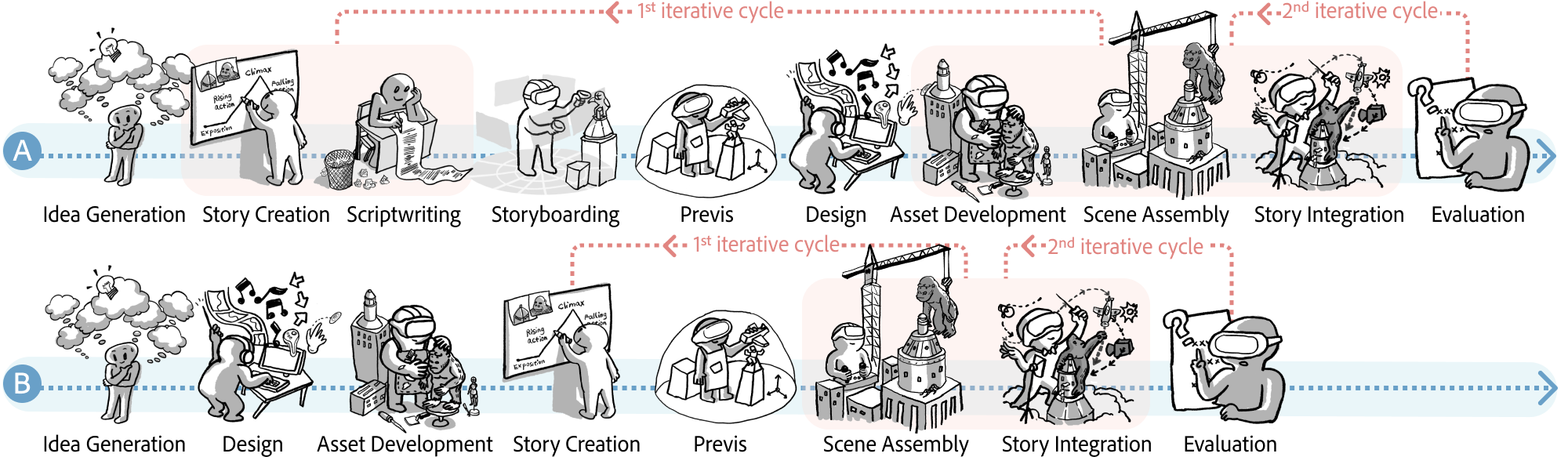}
    \caption{Typical workflows for crafting animated VR stories. (A) Story-driven workflow: prioritizing narrative coherence with visuals serving the story. (B) Visual-driven workflow: pursuing compelling visual effects that shape the story content.} 
    \label{fig:diverse_workflow}
    \Description{This figure illustrates typical story-driven (A) and visual-driven (B) workflows to craft animated VR stories.}
\end{figure*}

\wrap{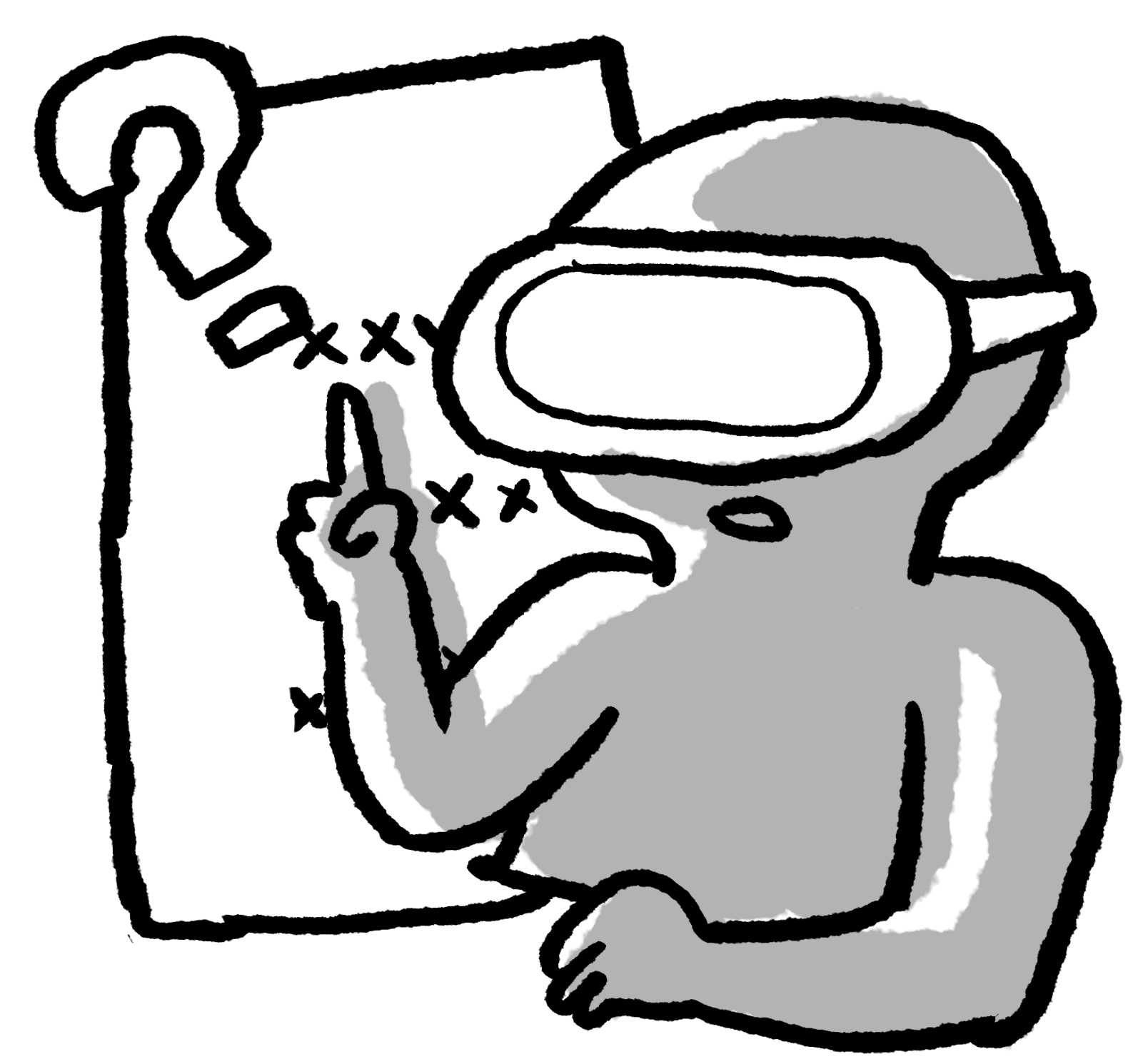}
\subsubsection{Evaluation}
\label{sec:evaluation}

Creators invite viewers to experience the final VR story and give feedback. Our interviewees primarily sought feedback on narrative comprehension, engagement, and user comfort through informal interviews or casual chats. 
They used two main ways to present their VR stories to viewers. First, many showcased their VR stories at exhibitions for walk-in visitors. Second, to reach a wider audience, they converted their VR stories into 360-degree panoramic videos, viewable on mobile phones and computers. Some also created 2D videos by recording their VR stories for online streaming platforms. 

\subsection{Story-Driven and Visual-Driven Workflows}
\label{sec:diverse_workflow}

Our interviewees embraced diverse workflows that connected the above ten stages, although some might omit some stages. Only \RR{nine, ten, eleven, and sixteen} interviewees were involved in scriptwriting, storyboarding, previs, and evaluation. The sequence in which these stages were carried out also varied. For instance, while some prioritized design before storyboarding and previs, others followed the reverse order. Two common iterative cycles emerged in their workflows. The first cycle occurred between story creation, scriptwriting, and production, while the second cycle appeared between production and evaluation.
Despite the diversity in these workflows, they can be classified as story-driven or visual-driven. Figure~\ref{fig:diverse_workflow} provides representative examples for each type. 

Most interviewees (16/21) adopted story-driven workflows (Fig.~\ref{fig:diverse_workflow}-A), usually in an order of the pre-production phase, production phase, and evaluation stage. They prioritized narrative coherence, ensuring that the visuals were designed to support and enhance the story. When certain visual elements proved difficult to implement, they often adjusted their designs to more practical or achievable solutions. For example, P3 changed his design from dreamlike to realistic to avoid additional workload.

In contrast, in visual-driven workflows (Fig.~\ref{fig:diverse_workflow}-B), interviewees (5/21; P1, P11, P14, P15, P16) exhibited a pronounced focus on visuals. They invested significant effort in exploration and experimentation. For example, P14 and P15 delved into rendering and shaders for stylized visuals, whereas P18 spent three months translating her 2D illustration visual style into VR using Quill. In visual-driven workflows, the stages in the pre-production and production phases were carried out alternately. Specifically, story creation followed the design and asset development stage, with the story often shaped by available assets, resulting in relatively simpler storylines.

\section{Creation Challenges}
\label{sec:challenges}

We identify seven challenges including a total of seventeen issues (\textbf{RQ2}), as listed in Table~\ref{tab:challenges}. Among them, eight issues echo the findings in previous empirical studies on general XR applications~\cite{krauss2021current, ashtari2020creating, krauss2022elements, liu2023challenges}. Therefore, we will only elaborate on the nine newly identified issues that exhibit certain uniqueness arising from the integration of story elements with VR and computer animation.

\begin{table*}
\setlength{\aboverulesep}{0.5pt}
    \setlength{\belowrulesep}{0.5pt}
\setlength{\tabcolsep}{6pt}
\definecolor{lightbluecolor}{HTML}{EDF5F7}

\caption{An overview of challenges in creating animated VR stories by creators, according to our interview studies. Items with a \colorbox{lightbluecolor}{colored background} are newly identified issues.}
\label{tab:challenges}
\Description{Overview of seven challenges.}
\begin{tabular}{@{}lp{5.1cm}m{7.6cm}p{3.2cm}@{}}
\toprule
{\#} & \multicolumn{1}{c}{\textbf{Challenge}}                                             & \multicolumn{1}{c}{\textbf{Issue}}                                                                                                                            & \multicolumn{1}{c@{}}{\textbf{Stage}}                                                     \\
\midrule
                &                           &  \cellcolor[HTML]{edf5f7}{C1-1: Lack of guidelines about balancing creators' narrative intent with viewers' autonomy}                                                                                            &  \\ %\hhline{~~-~} 
                                 &                                                                                      & \cellcolor[HTML]{edf5f7}{C1-2: Hard to understand VR story experiences and guidelines outside a VR environment}  &                                                                                           \\ %\cline{3-3}
                      \multirow{-4}{*}{\textbf{C1}}      &    \multirow{-4}{*}{ \parbox{5.1cm}{Insufficient VR storytelling guidelines}}                                                                              &  {C1-3: Limited guidelines for creators with different backgrounds to avoid applying incompatible knowledge (e.g., game design, 2D audio-visual language) }                                                                                    & \multirow{-5}{*}{\parbox{3.2cm}{Idea Generation, Story Creation, Scriptwriting, Storyboarding, Previs, Design}}                                                                                        \\ \hline
                    
             &               & \cellcolor[HTML]{edf5f7}{C2-1: Hard to describe and access multi-element plots based on story elements and desired outcomes}                                                                                     &                                     \\ %\hhline{~~-~}
                            \multirow{-3}{*}{\textbf{C2}}    &    \multirow{-3}{*}{\parbox{4.5cm}{Hard to find references for multi-element VR plots}}                                                                                     & {C2-2: Limited high-quality VR plots that align with creators' creative needs}                                                               &     \multirow{-2.5}{*}{\parbox{3.2cm}{Storyboarding, Previs, Design}}                                                                                        \\ \hline
                                       &       &  \cellcolor[HTML]{edf5f7}{C3-1: Difficult to plan and manage visual details under VR’s real-time performance constraints}                                                                                                   &                                             \\ %\hhline{~~-~}
                            \multirow{-3}{*}{\textbf{C3}}      &    \multirow{-3}{*}{\parbox{4.5cm}{Struggle to achieve satisfactory VR visuals for artistic expression}}                                                                                    &  \cellcolor[HTML]{edf5f7}{C3-2: Hard to achieve professional-grade visual quality with immersive tools alone}     &            \multirow{-3}{*}{Asset Development}                                                                                     \\ \hline
         &              & {C4-1: Hard to realize mismatches between creation mindsets and philosophies of DCCs and immersive tools}                                            &                                                   \\ %\hhline{~~-~}
                             \multirow{-3}{*}{\textbf{C4}}   &   \multirow{-3}{*}{\parbox{4.5cm}{Difficult to fluidly use non-immersive and immersive software}}                                                                               &  {C4-2: Inconvenient user interfaces and input interactions in DCCs and immersive tools}                                                                                    &     \multirow{-3}{*}{Asset Development}                                                                                  \\ \hline

             &    & \cellcolor[HTML]{edf5f7}{C5-1: Hard to plan and coordinate various story elements spatially and temporally}                                                                     &                               \\ %\hhline{~~-~}
                                &                                                                                     & \cellcolor[HTML]{edf5f7}{C5-2: Hard to switch between multiple narrative perspectives without confusing viewers about their roles}                                                                                     &                                                                                         \\ %\hhline{~~-~}
                                \multirow{-4.5}{*}{\textbf{C5}} &   \multirow{-4.5}{*}{\parbox{4.5cm}{Missing integrated building blocks for core VR story experiences}}                                                                               &  {C5-3: Hard to implement semantically-rich interactions related to story content}                                                                                     &     \multirow{-4.5}{*}{\parbox{2.5cm}{Scene Assembly, Story Integration}}                                                                                          \\ \hline
            &      & \cellcolor[HTML]{edf5f7}{C6-1: Uncertainty in the relationships between various design parameters,  emotions, and viewer comfort}                                 &                                   \\ %\hhline{~~-~}
                         \multirow{-3}{*}{\textbf{C6}}        &      \multirow{-3}{*}{\parbox{4.5cm}{Tedious parameter adjustment for optimal audience experience}}                                                                               &  {C6-2: Inefficient transition between a desktop and a VR environment to adjust various design factors based on firsthand VR experience}                                                         &    \multirow{-3}{*}{\parbox{2.5cm}{Scene Assembly, Story Integration}}        \\ \hline
                                      &    &  {C7-1: Unaware of key aspects during assessment}                                                                 &                               \\ %\hhline{~~-~}
                                &                                                                                     &  {C7-2: Hard to deal with the complex nature of individual audience experiences}                                                                                     &                                                                                         \\ %\hhline{~~-~}
                                \multirow{-4}{*}{\textbf{C7}} &   \multirow{-4}{*}{\parbox{4.5cm}{Lack data collection and analysis methods for evaluation}}                                                                              &  \cellcolor[HTML]{edf5f7}{C7-3: Insufficient support for collecting and analyzing viewers' watching behavior data}                                                                                      &     \multirow{-4.5}{*}{\parbox{2.5cm}{Evaluation}}                                                                                          \\
                                \bottomrule
\end{tabular}
\end{table*}

\subsection{C1: Insufficient VR Storytelling Guidelines}
\label{sec:challenge_guideline}

During the pre-production phase, our interviewees often found that existing storytelling guidelines were inadequate for addressing unique VR considerations and remained hard to understand.

\textbf{C1-1: Lack of guidelines about balancing creators' narrative intent with viewers' autonomy.}
Our interviewees were aware that viewers in VR have the autonomy to choose what to focus on and for how long. While some embraced this autonomy and told their stories indirectly (Sec.~\ref{sec:story_creation}), most interviewees experienced frustration as it often interfered with their narrative intent.
The first frustration is out-of-order exploration, where viewers encounter plot elements in a random order, disrupting planned revelations and suspense.
The second frustration is missing pivotal moments, where viewers focus on unintended aspects of the experience, weakening the intended emotional impact.
The third frustration is the disruption of narrative pacing, where interactive elements absorb viewers’ attention, causing temporal disconnects from the broader storyline.
Despite these frustrations, our interviewees were reluctant to excessively control or restrict viewers' autonomy. Approaches like limiting interactive elements or constraining movement and even gaze will undermine VR's unique advantages. However, they could not find relevant guidelines:

\q{In my story, you'll be Alice adventuring in Wonderland! You can explore freely, establish connections with the characters, and feel their emotions. However, I also have a predetermined storyline. Then, numerous questions emerge. How can I accommodate viewers' curiosity and self-exploration without risking narrative distraction? Can this curiosity be strategically used to enhance narrative engagement? How should I advance the storyline to maintain good narrative pacing, especially when interactive elements are more attractive?} (P13)

\textbf{C1-2: Hard to understand VR story experiences and guidelines outside a VR environment.}
Our interviewees indicated that the prevailing formats (e.g., text, images, or videos) of VR storytelling guidelines lack intuitiveness. This difficulty applies not only to the crafting of overall VR story experiences but also to specific design aspects. Managing visual hierarchy and weight within a boundless 360-degree canvas is one such design aspect. For example, P19, who learned about using 3D perspective lines and vanishing points from an online 2D video, found it difficult to apply these concepts in VR to establish a consistent focal point from different angles. 
Additionally, newcomers who had not yet fully understood the viewers' experience in VR stories also reported difficulties in understanding certain pieces of advice.

\q{... I questioned my teacher's advice to mirror real life in my designs. Later, I realized that in immersive VR, too many unfamiliar elements could disrupt narrative engagement more than on 2D screens... floating without gravity once took me out of the story, making me wonder why it happened.} (P5)

\begin{figure*}[!tb]
    % \centering
    \includegraphics[width=\linewidth]{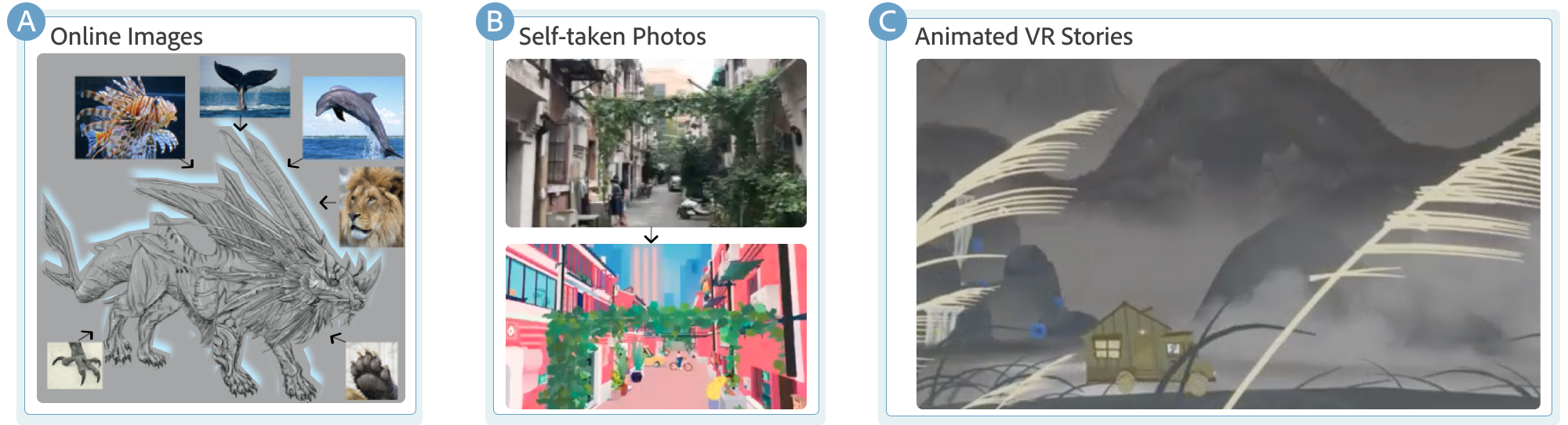}
    % \vspace{-8mm}
    \caption{Examples of different types of references. (A) Online images for the visual design of a character (\CR{\copyright Angela Cai}). (B) Self-taken photos to inform the design of a scene (\CR{\copyright AMAO}). (C) A multi-element plot from an animated VR story that showcases how to use spatial layering for aesthetic scenes and use a moving vehicle as symbolism tied to the story's theme (\CR{\copyright Xiaoka}).} 
    \label{fig:design_references}
    \Description{This figure provides examples of different types of references. Part A illustrates "Online Images," where various animal pictures (e.g., lion, whale) are combined to inspire a creature design. Part B shows "Self-taken Photos," where a real-world alleyway inspires a stylized 3D environment. Part C features "Animated VR Stories," showing the inspiration from a well-known VR story.}
\end{figure*}

\begin{figure*}[!tb]
    \centering
    \includegraphics[width=\linewidth]{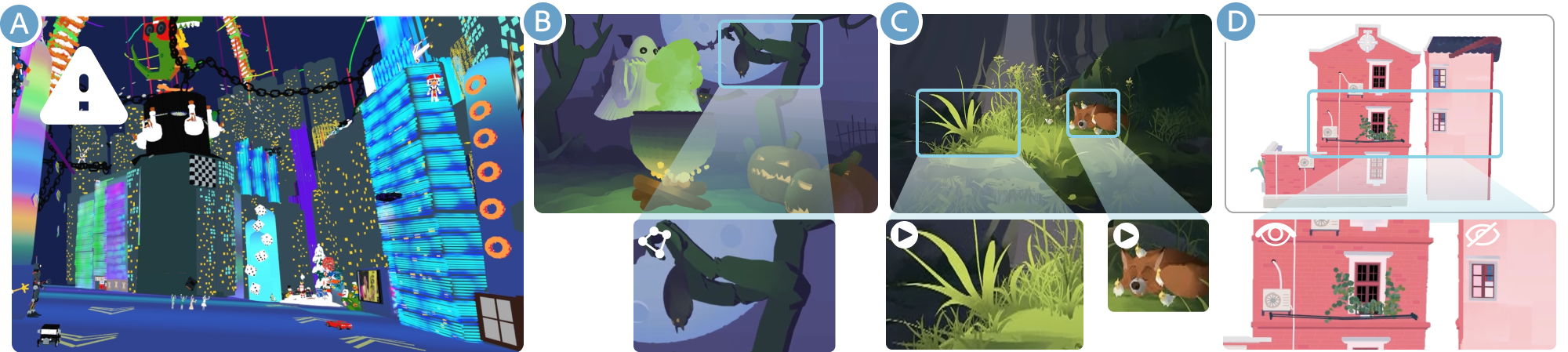}
    \caption{Examples of scenes that suffer from performance failure and relevant remedies. (A) A high-fidelity scene exceeding memory quota (\CR{\copyright Luna Han}). (B) Employing low-poly models to create scenes (\CR{\copyright Bigz}). (C) Listing assets to animate in advance to optimize capacity distribution (\CR{\copyright Bigz}). (D) Changing the level of details based on visibility (\CR{\copyright AMAO}).} 
    \label{fig:capacity_fidelity}
    \Description{This figure illustrates a scene suffering from performance failure and potential remedies. Part A shows an overloaded cityscape. Part B suggests using low-poly brushes to paint trees. Part C reduces complexity by spotlighting specific areas, such as a fox and grasses. Part D demonstrates controlling the details of different parts of a house based on visibility.}
\end{figure*}

\subsection{C2: Hard to Find References for Multi-Element VR Plots}
\label{sec:reference_plots}

Our interviewees underscored the critical role of references (Fig.~\ref{fig:design_references}) during the storyboarding, previs, and design stages. They particularly needed segments of animated VR stories to steer their design decisions to effectively interweave multiple visual, auditory, and interactive elements into engaging plots. For brevity, we refer to such segments as multi-element plots. 
As shown in Fig.~\ref{fig:design_references}-C, an interviewee took cues from the opening of \textit{Baba Yaga}~\cite{Babayaga}, imitating its spatial layering and lighting to craft aesthetic scenes, and mimicking its use of a vehicle as symbolism tied to her story's theme. However, finding multi-element plot examples is challenging.

\textbf{C2-1: Hard to describe and access multi-element plots based on story elements and desired outcomes.}
Our interviewees relied on describing their target plots to more experienced individuals to receive recommended references. They often began by describing their desired outcomes, including specific narrative effects, emotional impacts, and sensory experiences. However, they were not satisfied with simple and vague descriptions like \q{a symbolic plot evoking deep thoughts} for narrative effects (P14), \q{a plot giving viewers a sense of reverence} for emotional impacts (P10), or \q{a plot providing a cold atmosphere and physical sensation} for sensory experiences (P2). They wished to obtain more accurate recommendations by specifying preferences, such as viewers' position and viewpoint, the composition of multiple elements, and the integration of interactive elements like user choices. Articulating this combination of multiple elements and the resultant outcomes proved more difficult in animated VR stories than in films and animations. This complexity arose because, in VR, elements unfold not only visually and audibly but also spatially and interactively:

\q{If I ask for a plot where a crowd running around, I want to specify my preferences for camera movement and how the viewers engage within the story, such as their spatial relationship with the surroundings. Otherwise, my friend’s recommendations may not align with my expected outcome, but it is hard to express these specifics.} (P6)
\begin{figure*}[!tb]
    \centering
    \includegraphics[width=\linewidth]{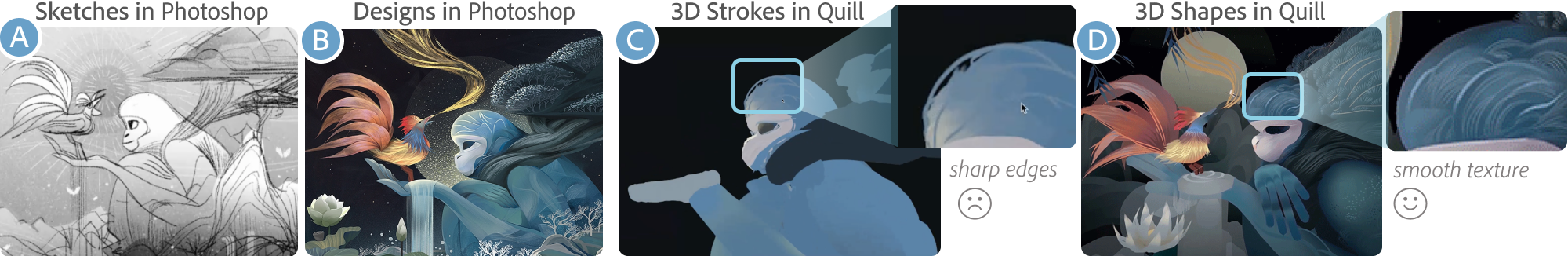}
    % \vspace{-8mm}
    \caption{Evolution of the main characters (\CR{\copyright Menghui}) from sketches (A), to designs (B), intermediate failed assets (C), and final successful assets (D). Building 3D models in Quill may require a mindset shift from using 3D strokes (C) to 3D shapes (D).}
    \label{fig:visual_style}
    \Description{This figure illustrates the evolution of the main characters in a participant’s story. Part A shows initial sketches created in Photoshop, focusing on rough outlines. Part B displays more refined designs in Photoshop, with detailed coloring and shading. Part C highlights the transition to 3D strokes in Quill, where characters exhibit sharp edges. Part D shows the final stage with 3D shapes in Quill, achieving smoother textures and a polished look for the characters.}
\end{figure*}

\begin{figure*}[!tb]
    \centering
     \includegraphics[width=\linewidth]{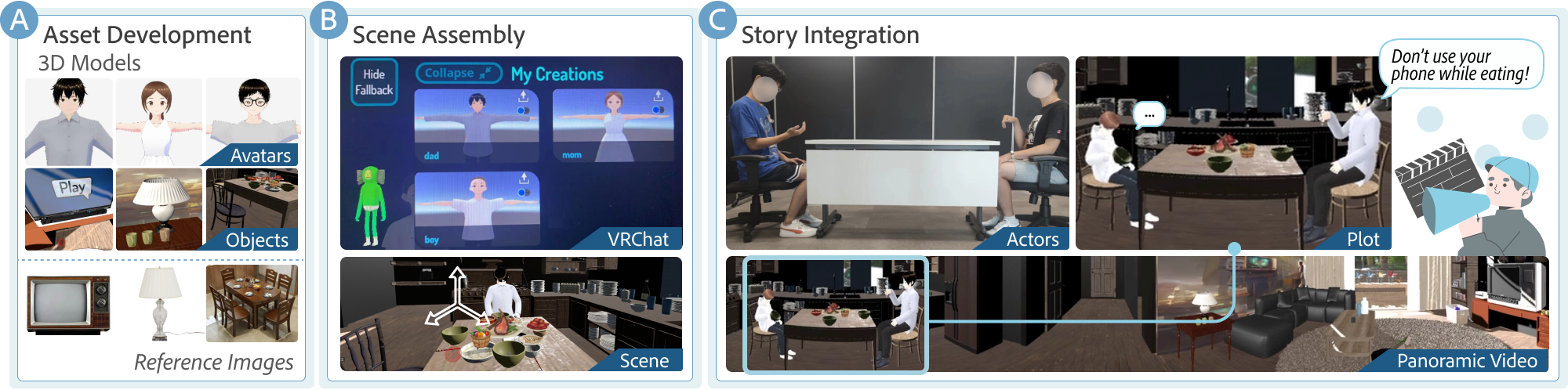}
    \caption{An innovative production strategy that leverages VRChat to simulate the live-action filmmaking process for spatial and temporal coordination (\CR{\copyright Hedi's team}). (A) Creating 3D models for story characters as VRChat avatars. (B) Uploading scenes and avatars to VRChat. (C) Simulating the shooting process in VRChat. With VRChat’s tracking capabilities, two creators control avatars and perform as actors based on the director’s instruction.} 
    \label{fig:vrchat}
    \Description{This figure illustrates an innovative production strategy using VRChat to simulate live-action filmmaking for an animated VR story. Part A showcases asset development, including references and 3D avatars. Part B focuses on scene assembly, where characters and environments are set up. Part C demonstrates story integration, using actors to simulate character behavior, developing the plot, and capturing panoramic video to create immersive scenes.}
\end{figure*}

\subsection{C3: Struggle to Achieve Satisfactory VR Visuals for Artistic Expression}
\label{sec:challenge_visuals}

In the asset development stage, interviewees always wanted better visual aesthetics and richer visual details, such as high texture resolution, vivid lighting and shadows, and lifelike character animations. These were important for their artistic expression and storytelling. However, despite their aspirations, they often struggled to achieve satisfactory visuals in their VR stories.

\textbf{C3-1: Difficult to plan and manage visual details under VR’s real-time performance constraints.}
Compared to pre-rendered films and animations, our interviewees encountered heightened conflicts between visual details and performance constraints. 
They often overlooked performance issues in pursuit of visual perfection and suffered from significant rework. In an extreme case, an interviewee, after dedicating 15 days to a scene (Fig.~\ref{fig:capacity_fidelity}-A), faced inadequate capacity and ultimately abandoned the project.
To mitigate rework, interviewees explored several solutions. 
For example, they might complete the entire story using low-poly models and brushes that consume less memory (Fig.~\ref{fig:capacity_fidelity}-B). If there was remaining capacity, a systematic upscale might follow. However, this approach might still compromise the visual quality and also lead to redundant work.
Another solution was to plan capacity distribution in each scene. For instance, an interviewee listed in advance which objects to animate (Fig.~\ref{fig:capacity_fidelity}-C), and another interviewee set priority levels for different parts of a scene and simplified those parts less visible to viewers (Fig.~\ref{fig:capacity_fidelity}-D). 
However, this approach hindered their desire for expressive freedom and was impractical: 

\q{Planning is often impractical because I tend to create what comes to my mind, especially when I am in a flow state. Besides, Quill allocates 1.5G memory per story, but its actual influence on my work is abstract. What can I draw and to what extent when I use a mixed set of brushes?} (P17)

\textbf{C3-2: Hard to achieve professional-grade visual quality with immersive tools alone.}
Our interviewees with backgrounds in 2D painting or illustration appreciated the immersive tools (e.g., Quill, TiltBrush) that enabled them to freely create animated VR assets and stories in a painterly manner (Fig.~\ref{fig:visual_style}). However, their creations were \q{far from meeting commercial use requirements} (P4), as these tools often fell short in achieving the desired visual quality. This shortfall was mainly due to \q{the limited range of brushes and the tools’ inability to produce rich color transitions and sophisticated lighting effects} (P21). When attempting to enhance the outputs from these immersive tools, our interviewees were pushed to reintegrate with \RR{digital content creation tools (DCCs)} and game engines. Such a shift confronted them with complex computer graphics:

\q{I initially chose Quill for its ease of use to avoid the steep learning curve of DCCs. Yet, when it came to re-rendering the rough surfaces in my Quill story to achieve a smoother appearance while preserving its warm aesthetic, I found myself needing to use shaders in Blender. The limitations of Quill brought me back to the difficulty I had hoped to escape.} (P14)

\subsection{C5: Missing Integrated Building Blocks for Core VR Story Experiences}
\label{sec:challenge_functionalities}

At the scene assembly and story integration stages, interviewees reported the absence of integrated, narrative-centric building blocks. For example, P7 stated, \q{For game development, I can find various building blocks such as integrated toolkits and prefabs. They provide a suite of features to ease gameplay and level implementation, but not many building blocks that are centered around VR narrative.}

\begin{figure*}[!tb]
    \centering
    \includegraphics[width=\linewidth]{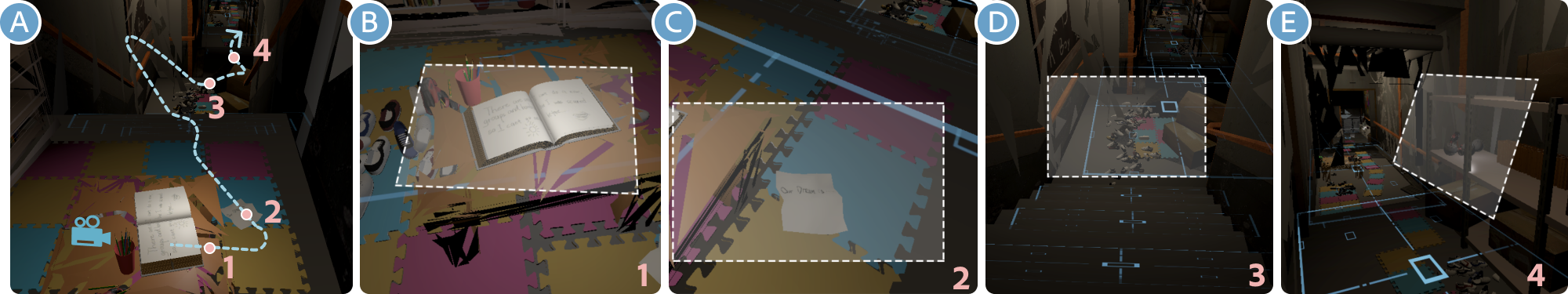}
    % \vspace{-8mm}
    \caption{An example of creator-intended exploration path (A) and POIs (B-E) (\CR{\copyright Julia's team}). These POIs could include (B) interactive objects, (C) narrative-related regions, and (D, E) areas to be noticed during teleportation.} 
    \label{fig:quantitative_evaluation_support}
    \Description{This figure illustrates a creator-intended exploration path (A) and POIs (B-E). Part A shows a path guiding the viewer through various key areas in a scene, marked by numbered points. Part B highlights the first POI, an open book placed on a colorful mat. Part C shows the second POI, a handwritten note. Part D focuses on the third POI, where the viewer's attention is drawn further down a hallway. Part E presents the fourth POI, a large framed picture.}
\end{figure*}

\textbf{C5-1: Hard to plan and coordinate various story elements spatially and temporally.}  
As shown in Fig.~\ref{fig:within_scene_integration}, our interviewees relied on tables and diagrams \q{as communication and planning materials before actual execution in creation software} to describe the status of characters, cameras, lights, and sounds for each plot. These methods were non-intuitive, offered limited granularity for capturing shorter moments, and required much time to compile such textual documents. 
Interestingly, we found that a team innovatively used VRChat to alleviate the difficulty of spatial and temporal coordination. They viewed VRChat as a tool, rather than a social VR platform, and incorporated it into their production (Fig.~\ref{fig:vrchat}). 
Specifically, they developed characters as VRChat avatars and assembled scenes in Unity before uploading them to VRChat. During story integration, the team entered their scene in VRChat, utilizing its head, hand, and full-body tracking capabilities to control avatars and act within virtual settings. This method mimicked live-action film shooting, with some members controlling different characters.
However, this method required collaboration among several people and additional tracking devices, which were not available for all creators. 
Thus, most interviewees assembled their scenes and weaved the scenes into a story manually using game engines, often complaining that the process was tedious and frustrating:

\q{I'm exhausted from manually aligning the multiple elements in multiple scenes, considering object placement and emotional impact. Though these elements are explicitly connected, I haven't yet found a way to leverage these connections to simplify the alignment.} (P4)

\textbf{C5-2: Hard to switch between multiple narrative perspectives without confusing viewers about their roles.}
Narrative perspectives determine the point of view from which viewers experience animated VR stories. Most interviewees utilized a fixed narrative perspective, either first-person or omniscient. Some interviewees (e.g., P6, P16, P20) favored multiple narrative perspectives. For instance, P20 frequently alternated between first-person and omniscient perspectives, so that \q{viewers can play different roles, ranging from being a candle actively engaging in interactions, to a passive observer gaining a comprehensive understanding of the unfolding events} (P20). However, they reported that achieving multiple perspectives required more than merely shifting camera positions; it necessitated careful preparation to ensure smooth transitions and maintain the viewers' sense of presence. This complexity was evident in P16’s attempt to connect disparate narrative branches by shifting perspectives between a refugee child and a journalist:

\q{To transition viewers into journalist roles, we used cues such as press badges, watching TV news, and mirrors for self-viewing. We also needed to connect these cues with the previous perspective of a refugee child. All these required additional work and risked disrupting a consistent and unbroken narrative experience, so we gave up.} (P16)

\subsection{C6: Tedious Parameter Adjustment for Optimal Audience Experience}
\label{sec:tedious_adjustment}
In the scene assembly and story integration stage, our interviewees needed to fine-tune various design factors and parameters to provide a better audience experience, which was tedious. The design factors included but were not limited to colors, lighting, 3D model sizes, camera movement, pacing, and spatial arrangements.

\textbf{C6-1: Uncertainty in the relationship between various design parameters, desired emotions, and viewer comfort.}
Our interviewees often sought to express strong feelings, such as excitement, fear, and exhilaration in their stories. To achieve this, they used camera movement, special effects, and light changes. However, they were often unsure whether and how their intended emotions could be conveyed well in VR with these means. This uncertainty prompted them to experiment with various solutions for conveying emotions and to tweak various parameters to avoid viewer discomfort, such as motion sickness. P19 shared her struggles:

\q{I wanted to convey tension, anxiety, and regret at different moments. I tried common types of camera movements for these emotions, such as quick cuts or tracking shots, but I felt motion sickness in VR... Is it possible to evoke emotions and create a certain atmosphere with cinematography in VR while preventing discomfort? If yes, how should I set parameters like speed, acceleration, and trajectories to resonate with distinct emotions?} (P19)

\subsection{C7: Lack Data Collection and Analysis Methods for Evaluation}

In the evaluation stage, though our interviewees wanted to enhance their VR stories based on audience feedback, they had difficulties in collecting and analyzing data.

\textbf{C7-3: Insufficient support for collecting and analyzing viewers’ watching behavior data.}
Our interviewees found it difficult to understand viewers' experiences based on verbal communication. Thus, they wanted support to collect and visualize viewers' behavior, such as their gaze and exploration trajectories. 
For example, an interviewee wanted to identify the differences between the viewers' actual exploration and her intended path (Fig.~\ref{fig:quantitative_evaluation_support}-A) and \RR{points of interest (POIs)} (Fig.~\ref{fig:quantitative_evaluation_support}-B-E) to refine the storylines. However, most interviewees found that such support was limited due to the complexity of VR stories:

\q{Guiding a viewer's gaze in VR spaces requires me to place various cues. I am not sure whether my visual cues are effective, so I want a function for data visualization. Besides, my story is continuous in both time and space, and the viewers' exploration is also continuous in time and space. How can I analyze them together?} (P16)

\section{Validation}
\label{sec:validation}

We invited both the original interviewees (N=21) and \RR{additional animated VR story creators (N=15)} to review the findings reported in Sec.
~\ref{sec:common_stages} and Sec.~\ref{sec:challenges}. The purpose was to assess the applicability, perceived importance, and perceived difficulty of the summarized stages and challenges.

\subsection{Setup}
To validate our findings, we designed two online questionnaires regarding the creation stages and challenges, respectively. Then we asked each respondent to complete the two online questionnaires. 

\textbf{Questionnaire for creation stages.}  The questionnaire described the ten stages and showed Fig.~\ref{fig:workflow_example_1} as a visual depiction. It then asked: (1) \q{Please select the three most important stages in completing an animated VR story.} (2) \q{Are there any stages in your creation of animated VR stories that these ten stages do not encompass?}

\begin{figure*}[!tb]
    \centering
    \includegraphics[width=\linewidth]{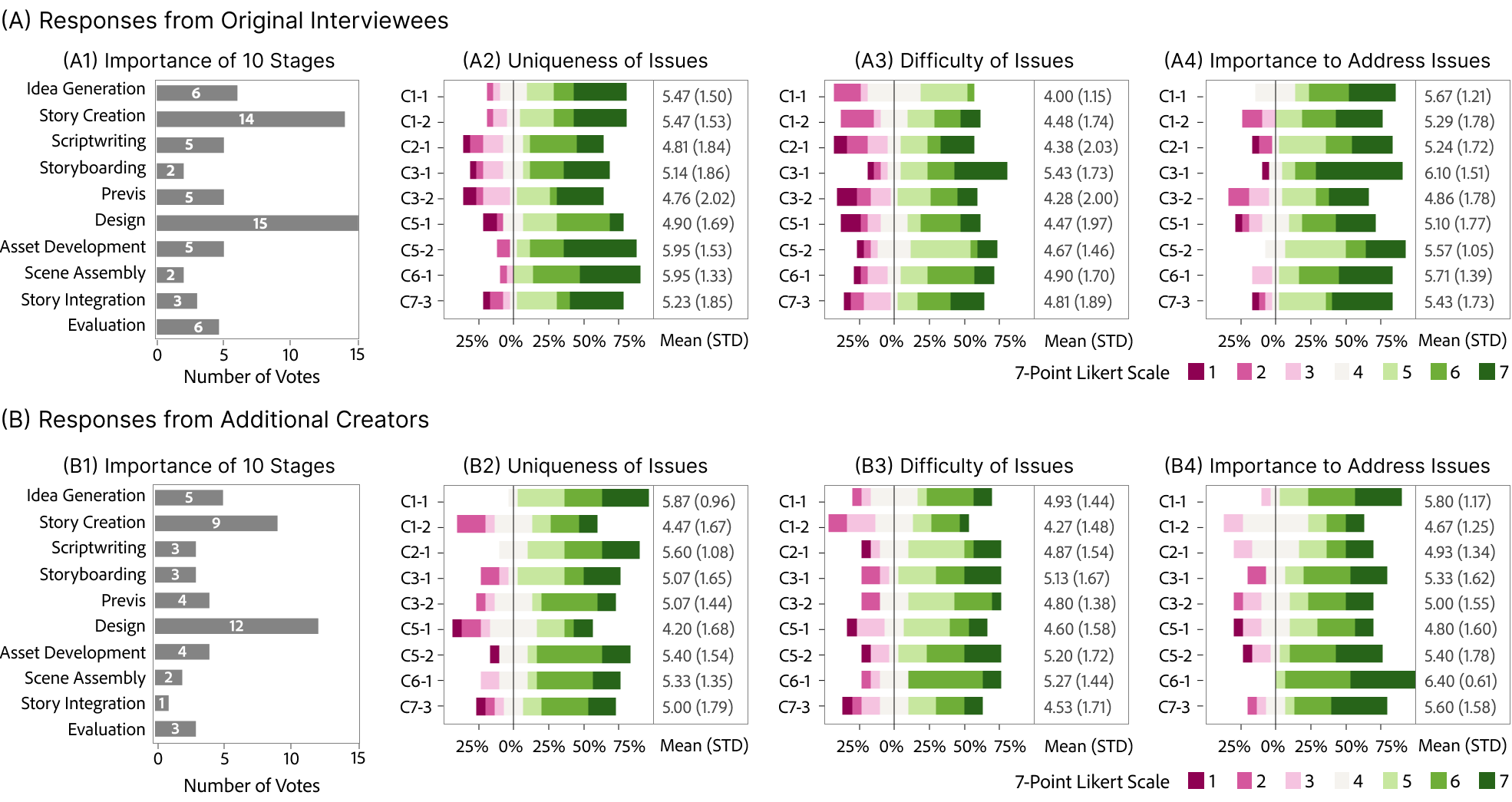}
    \vspace{-6mm}
    \caption{Summary responses from our 21 original interviewees (A1-A4) and \RR{15 additional creators (B1-B4)}, including the number of votes for each stage regarding the importance (A1, \RR{B1}), subjective ratings on the uniqueness (A2, \RR{B2}) and difficulty (A3, \RR{B3}) of newly identified issues, and importance (A4, \RR{B4}) to address the issues. 1: Not unique/difficult/important at all. 7: Extremely unique/difficult/important.} 
    \label{fig:interviewee_rating}
    \Description{This figure has two parts, summarizing validation results for the importance of stages and challenges from our original interviewees and additional creators. Part A is the results from our original interviewees, and Part B is the results from the additional creators. From left to right, each part consists of votes for important stages, uniqueness of issues, difficulty of issues, and the importance of addressing these issues.}
\end{figure*}

\textbf{Questionnaire for creation challenges.}
The questionnaire had two parts. The first part described each newly identified issue in turn, followed by three questions rated on a 7-point Likert scale: (1) \q{Compared to other forms of stories, to what extent is this issue unique to animated VR stories?} (2) \q{How much difficulty does this issue pose during your VR story creation?} (3) \q{Assuming we want to offer better creativity support, how important is it to solve this issue?} 
The second part presented all 17 identified issues and asked whether any additional issues were missed.

\RR{\textbf{Respondents.}}
\RR{We had a total of 36 respondents, consisting of 21 original interviewees (Sec.~\ref{sec:interview_recruitment}) and 15 additional creators. The interviewees helped verify our interpretations of their input, while the additional creators could further assess the comprehensiveness of our findings.
To recruit additional creators, we distributed our questionnaires to several VR creators' online communities and art schools.
After excluding three respondents who had not completed one animated VR story, we had 15 valid additional creators (aged 22-35; 7 females and 8 males). All creators were from China, except one from Japan.
Similar to our interviewees, all creators had formal art training and experience in different artistic fields such as digital media arts and animation. Their experience in non-VR art averaged 4.9 (min=2, max=9) years and their VR creation experience averaged 2.2 (min=1, max=5) years.
They had created an average of 2 (min=1, max=5) VR stories, with the creation time per story averaging 2.7 (min=1, max=6) months.}

\subsection{Analysis and Results}
\RR{The votes and ratings were generally consistent between the original interviewees and additional creators, with both groups agreeing on our findings. The detailed analysis and results are as follows.}

\textbf{Importance of stages.} Figure~\ref{fig:interviewee_rating} shows the number of votes for each stage selected as the top three important stages by both our original interviewees (Fig.~\ref{fig:interviewee_rating}-A1) and \RR{additional creators (Fig.~\ref{fig:interviewee_rating}-B1)}.
\RR{Both groups consistently identified story creation (14/21 original, 9/15 additional) and design (15/21 original, 12/15 additional) as the most crucial stages.} This consensus aligns with our findings on story-driven and visual-driven workflows (Sec.~\ref{sec:diverse_workflow}), where the story and design determine the direction of subsequent stages and outline the final output content.
\RR{Scene assembly (2/21 original, 2/15 additional) and story integration (3/21 original, 1/15 additional) received limited votes across both groups}, likely because the respondents prioritized stages demanding creativity. 
They viewed scene assembly and story integration as basic and manual stages of production, where the quality of results depends on human labor and skill level rather than creativity.
Nevertheless, all stages were considered important by at least one respondent in each group. 

\textbf{Uniqueness, difficulty, and importance of challenges.}
All the nine issues received mean ratings above 4 on a 7-point Likert scale across all three aspects from the interviewees (Fig~\ref{fig:interviewee_rating}-A2, A3, A4) and \RR{creators (Fig~\ref{fig:interviewee_rating}-B2, B3, B4), suggesting that both groups recognized the nine issues as unique, difficult, and important.} \RR{Furthermore, we performed Mann-Whitney U tests to compare the ratings of the two groups for each issue across the three aspects. Under a significance level of 0.05, no significant differences were found between the two groups for any issue. While there were variations in mean ratings between groups, such variations were expected and natural, as individual experiences could influence subjective ratings. Based on these non-significant test results, we proceeded to aggregate the raw ratings from both groups. The aggregated mean ratings show that the three most unique issues were C5-2 (mean=5.72), C6-1 (mean=5.69), and C1-1 (mean=5.64); the three most difficult issues were C3-1 (mean=5.31), C6-1 (mean=5.06), and C5-2 (mean=4.89); the three most important issues were C6-1 (mean=6.00), C3-1 (mean=5.78), and C1-1 (mean=5.72).} 
\RR{Interestingly, C1-1, C5-2, and C6-1 all relate to viewers, focusing on viewer autonomy, perspective management, and comfort. The results highlight that respondents were particularly concerned with effectively delivering their narratives while accommodating the unique affordances of VR, where viewers possess unprecedented agency and presence within the story space.}

\textbf{Comprehensiveness of stages and challenges.} 
None of the interviewees or \RR{additional creators} reported creation stages beyond our findings, indicating that our ten stages can describe the animated VR story creation process well. \RR{Respondents valued our clear stage identification, with one noting \q{my stages often overlap and blend together, so it's helpful to see them clearly laid out in this structured way, which will help organize my workflow in future projects.}}
In terms of challenges, our interviewees found that our summary reflected their input well and did not report any new challenges. \RR{Most additional creators also agreed that the challenges encompassed the obstacles they encountered. One creator suggested the challenge of \q{managing the intensity of VR experiences to avoid motion sickness or fatigue}, which fell under C6-1. Another creator mentioned \q{struggling to optimize VR story content for VR devices with varying specifications}, which aligned with C6-2. Therefore, no distinct challenges were identified beyond our findings.}
\section{Discussion}

This section compares animated VR stories with general XR applications~\cite{ashtari2020creating, krauss2022elements, liu2023challenges, krauss2021current}, proposes research opportunities, and acknowledges study limitations.

\subsection{\RR{Comparison between Animated VR Stories and General XR Applications}}

Recent studies have identified and discussed four phases to create general XR applications: requirement collection~\cite{borsting2022software}, designing and prototyping~\cite{ashtari2020creating, krauss2022elements}, implementation~\cite{ashtari2020creating, krauss2021current}, and testing~\cite{ashtari2020creating, liu2023challenges}. The ten stages identified in our study share some similar goals and activities with these phases but also exhibit differences.

Specifically, the six stages, from idea generation to design, are regarded as the pre-production phase of creating an animated VR story. Similar to requirement collection, designing, and prototyping in general XR creation, these stages focus on conceptualization and planning, gradually refining elements from low to medium fidelity~\cite{ashtari2020creating, krauss2022elements}. 
However, they also differ significantly.
For example, general XR applications, which are often designed for functional purposes~\cite{ashtari2020creating} such as training and rehabilitation, necessitate a strong focus on end-user needs during requirement collection~\cite{borsting2022software}. Conversely, our results show that VR story creators often commence from a place of personal expression, curating content based on the stories they desire to narrate (Sec.~\ref{sec:idea_generation}). 
This unique narrative focus leads to distinct issues related to balancing narrative intent with view autonomy (\textbf{C1-1}), understanding VR story experiences (\textbf{C1-2}), and describing multi-element plots (\textbf{C2-1}). 
These issues delve deeper into the core of VR stories than findings about insufficient general guidelines~\cite{ashtari2020creating} and a lack of general references~\cite{ashtari2020creating, krauss2021current} in prior XR research.

Asset development, scene assembly, and story integration are stages in the production phase. 
Similar to the implementation phase of general XR applications~\cite{krauss2021current}, the production phase also utilizes many tools (e.g., Unity, Blender) to transform designs into immersive VR experiences. 
However, there are several differences.
First, while general XR creators prioritize functionalities and interactions~\cite{ashtari2020creating, krauss2022elements}, VR story creators make more efforts to achieve visual quality and aesthetics (Sec.~\ref{sec:asset_development}). Thus, we highlight issues related to satisfactory visuals and artistic expressions (\textbf{C3-1}, \textbf{C3-2}), which are not emphasized in previous XR studies~\cite{ashtari2020creating, krauss2021current, krauss2022elements}. 
Second, VR story creators concentrate on specific and core VR story experiences, leading to unique issues in coordinating multiple story elements (\textbf{C5-1}) and switching between multiple narrative perspectives (\textbf{C5-2}).
Third, our results emphasize creators' efforts in \RR{adjusting design factors (e.g., colors, sizes, camera movement, and lighting configurations)} to enhance emotional impact (\textbf{C6-1}), \RR{while the emotional impact in VR has received limited attention in previous research~\cite{ashtari2020creating, krauss2021current, krauss2022elements}.}

The testing phase of general XR applications intersects with our evaluation stage, both aiming to verify that the final outputs function as anticipated. While the testing phase for XR applications encompasses various methods~\cite{liu2023challenges, borsting2022software}, including unit tests, integration tests, system tests, and user testing, animated VR story creators do not perform these testing methods. Once their VR story experience operates seamlessly, they proceed to informal user evaluation to gather viewers' feedback. 

\subsection{\RR{Future Research Opportunities}}
\label{sec:opportunities}

Based on our findings, we suggest several opportunities for future HCI research to support animated VR story creation. 

\textbf{Investigate narrative intent and view autonomy beyond guiding viewer attention.}
Our results reveal the need for guidelines about effectively conveying creators' intended messages and emotions in VR while preserving viewer autonomy (\textbf{C1-1}). Most existing HCI studies~\cite{rothe2019guidance, schmitz2020directing} equate this objective with directing viewers' attention to where critical plots happen, investigating various guidance mechanisms with audiovisual cues. Although these studies can mitigate frustrations such as out-of-order exploration and missing pivotal moments, they mainly prioritize creators’ narrative intent and proactively shape viewer autonomy. However, our results show that creators are interested in understanding and accommodating viewer autonomy, seeking to harmoniously integrate it into their storylines. In response to this interest, further studies beyond guidance mechanisms are needed. 
A recent HCI study~\cite{aitamurto2021fomo} demonstrates such an example that suggests focusing on viewers’ joy of missing out besides fear of missing out in VR storytelling. 
Future research can draw inspiration from various well-established creative fields, such as theatrical performance~\cite{pope2017geometry, gupta2020roleplaying}, 2D/3D game storytelling~\cite{poretski2022gameLearn}, and data visualizations~\cite{li2023geocamera}, to investigate whether and how existing theories and practices can be adapted to integrate viewer autonomy into VR stories featuring predefined storylines, environmental storytelling, or branch storytelling.

\textbf{Accommodate viewer autonomy in authoring tools.}
To further facilitate the balance between narrative intent and viewer autonomy (\textbf{C1-1}), we recommend that future authoring tools provide explicit features that take audience autonomy into account, particularly during the stages of story creation, previs, and design.
Our results indicate that creators' attempts to guess viewer exploration behaviors during the storyboarding or previs stage have not been notably effective, as evidenced by reported frustrations (Sec.~\ref{sec:challenge_guideline}).
Regrettably, most tools mentioned in Sec.~\ref{sec:related_authoring_story} prioritize assisting creators in designing and prototyping their envisioned content, while overlooking viewer autonomy.
Future authoring tools may embrace a paradigm shift to enhance consideration for viewer autonomy.
Some systems~\cite{nebeling2020xrdirector, rajaram2023reframe} have demonstrated such possibilities.
For example, REFRAME~\cite{rajaram2023reframe} allows creators to anticipate and address potential threats in the early design stage from a user's perspective by personifying various threats as characters in storyboards.
Besides, our results also indicate that there are already some guidelines related to viewer autonomy, but due to not being in VR environments, creators cannot understand them (\textbf{C1-2}), such as using 3D perspective lines in online videos or those guidance mechanisms in academic papers~\cite{rothe2019guidance, schmitz2020directing}. Thus, future authoring tools can start by examining these existing guidelines and investigating innovative features accordingly. 

\textbf{Explore effective representations for describing multi-element plots.}
Our findings highlight the complexity of describing multi-element plots in animated VR stories (\textbf{C2-1}). These plots encompass various aspects, including story elements (e.g., text, narration, camera), themes (e.g., breaking ice, time-travel), emotions (e.g., excitement, tension), spatial configurations (e.g., multi-layering), temporal dynamics (e.g., quick or slow pace), and even viewers' potential viewpoints.
To address this complexity, effective representations are essential. They will not only facilitate communication among creators but also underpin future authoring tools for collecting, managing, retrieving, and using multi-element plot references. 
Several recent HCI studies have proposed some representations~\cite{liu2019view, mahadevan2023tesseract} and showcased potential usages~\cite{stemasov2023sampling, mahadevan2023tesseract} for VR assets and environments.
For example, Tesseract~\cite{mahadevan2023tesseract} employs worlds in miniature to query and locate spatial design moments. 
Stemasov~\etal~\cite{stemasov2023sampling} suggested breaking down and remixing virtual objects by attributes such as colors or motion paths.
However, they do not target more complex animated VR stories, where story elements unfold spatially, temporally, and interactively. Future research may aim to propose representations capable of tracking dynamic elements, capturing interactive storytelling mechanisms, and facilitating spatial and temporal annotations and linkages among elements. Future research may further consider how to design representations of multi-element plots to help creators uncover complex relationships between elements and their contributions to high-level goals like emotional elicitation and narrative comprehension.

\textbf{Investigate context-sensitive algorithms and tools for optimizing visuals.}
Our results show that animated VR story creators face heightened conflicts between their pursuit of satisfactory visuals and performance constraints (\textbf{C3-1}). They also find it hard to achieve certain visual quality because of the inability of existing immersive tools such as Quill (\textbf{C3-2}). 
Solving these issues requires combining HCI research and computer graphics algorithms. 
One key area of exploration is how viewers perceive visual flaws in VR storytelling. This calls for HCI studies to investigate the impact of degraded texture quality or model fidelity across different visual elements, such as characters, props, and environments, and how these downgrades affect the narrative experience.
Informed by these insights, there is potential for the development of computer graphics algorithms that dynamically adjust visual fidelity in response to narrative demands. For example, while recent advancements in deep learning methods~\cite{li2023generative, hirzle2023xrmeetai} have enhanced 3D asset development processes like modeling and texturing, they often do not consider the specific performance optimization needs of VR devices. Embedding these insights into these methods could result in algorithms that selectively adjust visual fidelity, preserving high-quality visuals in the most impactful areas while optimizing resource use in less critical areas, thereby enhancing the narrative without straining VR resources.
Furthermore, these algorithms could be leveraged to enhance animated VR authoring tools, guiding creators more intuitively through the optimization process. Such tools could provide context-sensitive suggestions and adjustments. For instance, in dialogue-heavy scenes, optimizations might focus on enhancing character facial expressions while simplifying background elements. Conversely, in action-intensive scenes, the emphasis could be on maintaining fluid motion and visual coherence.

\textbf{Utilize connections between story elements for coordination.}
Our results indicate that the process of coordinating multiple elements spatially and temporally is cumbersome (\textbf{C5-1}). Although creators often identify explicit connections (e.g., semantic links, common social activities, and spatial relationships) between story elements, they lack effective tools to manifest these connections. Future research may first figure out what connections are common in animated VR stories. This exploration might draw inspiration from AR story authoring tools~\cite{li2022ARSemantics, li2023humanscene, li2024anicraft, tong2024vistellar}, which utilize semantic links~\cite{li2022ARSemantics} between virtual objects to distribute story events and employ common social behaviors~\cite{li2023humanscene} to generate character activities. However, these AR-based approaches might not fully address the unique demands of VR, especially in imagined scenes and activities. 
Therefore, further investigation is necessary to adapt and extend these methods for effective VR storytelling.

\textbf{Offer customizable modules for switching between multiple narrative perspectives.}
Our results reveal that VR story creators may want viewers to engage with their stories from different narrative perspectives, but they worry about the risk of disrupting consistent story experiences due to the difficulty in effectively switching between multiple perspectives (\textbf{C5-2}).
To address this issue, future HCI studies need to first understand whether and how viewers can perceive their roles when narrative perspectives change. However, current studies~\cite{bindman2018bunny, gupta2020roleplaying} indicate that viewers struggle to identify with a single role within a consistent perspective, let alone multiple roles across varying perspectives. This significant research gap suggests a need for collaboration between HCI experts and VR storytellers. By working together, they can develop compelling VR stories that incorporate diverse perspectives, thereby setting the stage for more focused experiments. These experiments could investigate which switching mechanisms can alleviate viewers' confusion and enhance their perception of new roles. Subsequently, effective switching mechanisms can guide the design of authoring tools. For instance, these tools could offer templates or modules for common perspective shifts, such as transitioning from a participant role to an observer role, to enable quick customization.

\textbf{Provide computational design support for enhancing emotional impact.}
Our results show that VR story creators need to adjust various design factors, aiming to enhance emotions and viewer comfort. 
This continuous tweaking makes the creation process less pleasant and often confuses creators due to the unclear quantitative relationships between design factors, emotions, and viewer comfort in animated VR stories (\textbf{C6-1}). While many HCI studies have explored how different factors, like spatial complexity~\cite{shin2022spatialcomplexity, shin2019anyroom}, jump cuts~\cite{zhang2024jump}, and perceived roles~\cite{bindman2018bunny}, influence audience experiences, only limited studies~\cite{serrano2017movie, hu2019reducing, xie2024emordle} have delved into modeling these relationships quantitatively. For instance, Hu~\etal~\cite{hu2019reducing} investigated how camera speed, acceleration, and scene depth impact perceived discomfort. To free creators from manual and trial-and-error adjustments by providing computational design support, future research may focus on modeling the relationships among design factors, emotions, and various devices (e.g., desktop and VR environments) in the context of animated VR stories. This could build on existing VR datasets~\cite{xue2021ceap,sitzmann2018saliency}, self-collected datasets~\cite{hu2019reducing}, or generated datasets~\cite{yuan2024generating}. Once these models encompass the myriad of design factors, emotions, and viewer comfort, upcoming VR creation systems could introduce semi-automatic or automatic algorithms (e.g.,~\cite{cha2020enhanced, pavel2017shot, hartmann2020effect}). Such advancements would aid creators in tasks like modulating lighting to convey emotions like fear while ensuring viewer comfort.

\textbf{Design data visualization tools for revising storylines with semantics.}
Our findings indicate that creators lack adequate support for gathering and analyzing data on viewer behaviors, which is crucial for refining storylines and visual cues in their stories (\textbf{C7-3}). Data visualization emerges as a potential solution to this issue. There are existing tools~\cite{nebeling2020mrat, buschel2021miria, hubenschmid2022relive} designed for general XR applications that aid in data collection and analysis. For instance, MIRIA~\cite{buschel2021miria} facilitates in-situ analysis of spatial and temporal interaction, while Relive~\cite{hubenschmid2022relive} enhances data analysis through visualization on desktops and VR headsets. Nonetheless, these tools primarily address static scenes, whereas animated VR stories evolve in both space and time. Moreover, their primary objective is to identify user behavior patterns and answer specific research questions, such as space utilization and social interaction patterns. They do not focus on aiding VR storytellers in understanding viewer engagement, narrative comprehension, or the effectiveness of their storylines. These more abstract goals, involving a semantic understanding of storylines, necessitate more specialized data visualization and analytic tools.

\subsection{Limitations}

Firstly, our interviewees were neither XR experts nor novice XR creators. Most had formal art training with experience in 2D/3D animation and game development, often using game engines. Their backgrounds might influence their focus on the pre-production phase requiring more creativity, rather than production. Consequently, our findings primarily apply to creators with similar experience levels. Future research may include participants like non-professional artists or hobbyists without art and XR backgrounds, to understand if their challenges and needs differ.

Secondly, our interviewees usually collaborated with professional audio production teams or utilized existing online resources to design and develop audio elements. Consequently, their insights into audio aspects were limited and not thoroughly explored in our study. Future research may engage creators specialized in audio design, particularly those experienced in storytelling through sound, to uncover innovations in audio design for VR stories.

Thirdly, given the recent availability of VR devices and creation software, our interviewees were predominantly between 22 and 33 years old. This may have excluded seasoned creators, who might offer unique insights and deeper creative needs from their broader experience in non-VR fields. Future empirical studies may consider including seasoned creators, such as filmmakers or animators who possess a desire to delve into animated VR stories but have not had the opportunity to do so.

\section{Conclusion}

Our study delved deeply into the creation processes and challenges associated with crafting animated VR stories. We conducted semi-structured interviews with 21 animated VR story creators. Through this, we identified ten stages that creators typically go through when creating animated VR stories. The inclusion and order of these stages can vary, leading to diverse workflows, which can be categorized as story-driven or visual-driven. Additionally, we highlighted nine unique issues that VR story creators encounter. Our findings complement existing research by offering a richer, more granular perspective on animated VR story creation. Based on the findings, we discuss several future research directions in supporting animated VR story creation.

\begin{acks}
The authors sincerely thank the interviewees and survey respondents for generously sharing their time, insights, and creativity. We are also deeply grateful to the reviewers for their constructive feedback.
This work was partially supported by Hong Kong Research Grants Council under the Areas of Excellence Scheme grant AoE/P-601/23-N and RGC GRF 16214623.
\end{acks}
% \newpage
\balance
\bibliographystyle{ACM-Reference-Format}
\bibliography{main.bib}

\appendix
\section{\RR{Pre-Interview Questionnaire}}
\label{appendix:questionnaire}

\RR{The pre-interview questionnaire asked the interviewees to share the information of their favorite VR story projects through the following questions:}

\RR{\begin{itemize}
    \item What was your favorite self-created animated VR story about?
    \item How many months did it take to complete your favorite animated VR story?
    \item What motivated your choice of VR as the medium for your favorite animated VR story?
    \item Did your favorite animated VR story have a target audience? If so, who was your target audience?
    \item What is the online link to your favorite animated VR story? Input “None” if not applicable. 
\end{itemize}
}

\section{\RR{Interview Guide}}
\label{appendix:guide}

\RR{The following questions served as a semi-structured interview guide to gather insights addressing the two research questions on creation processes (RQ1) and challenges (RQ2).}

\subsection{\RR{Creation Processes (\textbf{RQ1})}}

\RR{\begin{itemize}
    \item Thinking about your favorite VR story, what were the main stages you went through from scratch to final completion? Please describe the stages chronologically and illustrate with intermediate results by sharing your screen.
    \item For each stage you mentioned:
    \begin{itemize}
        \item What were your primary objectives in this stage?
        \item What specific tasks did you perform in this stage?
        \item What were the inputs and outputs of this stage? 
        \item What tools and software did you use in this stage?
        \item How did using VR and computer animation technologies make this stage different from other storytelling forms (e.g., 2D/3D animation, 2D/3D games, micro-films)?
        \item Compared to other storytelling forms, what unique considerations did you have in this stage for your animated VR story?
    \end{itemize}
    \item Regarding the overall workflow you described:
    \begin{itemize}
        \item How did you decide on the sequence of the stages? 
        \item Were there times you needed to revisit earlier stages? What prompted this?
        \item Did you work on multiple stages in parallel? What led to that decision?
    \end{itemize}
\end{itemize}
}

\subsection{\RR{Challenges (\textbf{RQ2})}}
\RR{For each stage that you mentioned:}

\RR{\begin{itemize}
    \item What unique difficulties did you encounter when integrating multiple story elements within VR and computer animation constraints? Please provide concrete examples.
    \item Compared to other storytelling forms, which tasks were particularly time-consuming? Why?
    \item Compared to other storytelling forms, what features or ideas did you initially plan but couldn't implement in the final VR story? Why?
    \item Compared to other storytelling forms, what were your most frustrating moments? Why?
    \item Earlier you mentioned [specific difference or unique consideration] compared to other storytelling forms. What challenges emerged from this?
\end{itemize}
}

\section{Project Details of Interviewees}
\label{appendix:project}
Table \ref{tab:project_details} summarizes the project details mentioned by our participants during the interview study.
\renewcommand{\arraystretch}{1.25}
\begin{table*}[ht]
\setlength{\aboverulesep}{0.5pt}
\setlength{\belowrulesep}{0.5pt}
\setlength{\tabcolsep}{8pt}
\caption{A summary of animated VR stories mentioned by interviewees. ID: Interviewee ID; \#Month: Number of months it took to finish the project; Software: Main authoring software employed in the creation process; Set-up of Animated VR stories: Self-reported story description regarding the viewers' experience.}
\Description{A summary of animated VR stories mentioned by interviewees.}
\label{tab:project_details}
\begin{tabular}{m{0.9cm}m{0.7cm}m{3cm}m{11cm}}
\toprule

\textbf{ID} & \textbf{\#Month} & \textbf{Software} & \textbf{Set-up of Animated VR Stories} \\ 
\midrule
P1 & 6 & Maya, UE & In a fantasy world inhabited by mythical creatures, viewers witness a grand homage ceremony to God by these beings. \\
\arrayrulecolor{lighttablerowcolor}\cline{1-4}
P2 & 1 & Maya, UE & Navigating five intertwined landscapes, viewers develop their own stories, delving into the unique, surpassing beauty of VR. \\
\cline{1-4}
P3 & 2 & C4D & A young man's tale of anxiety-induced escape from reality unfolds, inviting immersion in his ideal world and prompting reflection on escapism. \\
\cline{1-4}
P4 & 1 & Quill, Unity & Viewers accompany a caravan traversing the Silk Road, encountering diverse landscapes and cultures, and absorbing the mystique of the historic trade route. \\
\cline{1-4}
P5 & 1 & 3ds Max, Unity & A story sweeps viewers back in time to a local heritage site, where they relive the heroic deeds of a historical figure, absorbing the profound essence of history.\\
\cline{1-4}
P6 & 4 & Maya, UE & Acting as an ordinary citizen, viewers experience the rapid changes of a city through different historical stages, feeling the pulse of urban progress. \\
\cline{1-4}
P7 & 2 & TouchDesigner, Maya, UE & Viewers follow a game character on a bizarre adventure, where the character has unusual behaviors, leading them to re-explore the game in a new perspective. \\
\cline{1-4}
P8 & 6 & Maya, UE & Viewers encounter strange events in a virtual driving adventure, where the journey in VR attempts to blend cinematic techniques with dramatic conflicts. \\
\cline{1-4}
P9 & 5 & Maya, UE & As companions to a young protagonist, viewers learn to ski, navigate failures, and triumph over danger, ultimately mastering skiing. \\
\cline{1-4}
P10 & 3 & Maya, UE & As time-space explorers, viewers experience technology's limits, confront intricate situations, and grapple with moral dilemmas due to rapid technical advancement. \\
\cline{1-4}
P11 & 1 & Tilt Brush, Quill & Viewers find themselves at a lion dance competition, watching two teams fiercely compete, feeling the festive joy and fantasy of the New Year. \\
\cline{1-4}
P12 & 4 & VRChat, Maya, Unity & Viewers watch a child and father initially struggle without maternal love, but slowly reach understanding, revealing the depth of family bonds. \\
\cline{1-4}
P13 & 5 & Maya, UE & As a girl, viewers drift into a subway wonderland, meet magical beings, and uncover self-worth and true dreams. \\
\cline{1-4}
P14 & 3 & Blender & Viewers track a woman in her fifties navigating family troubles and bravely embarking on a solo road trip to start a new life. \\
\cline{1-4}
P15 & 6 & Quill, Blender & In a vibrant city, viewers uncover a tale of two sisters. Their interwoven destinies reveal a labyrinth of emotions and secrets concealed beneath the city's exterior. \\
\cline{1-4}
P16 & 2 & 3ds Max, UE & Viewers can embody a reporter, refugee child, or doctor, directly engaging in an anti-war narrative and experiencing the harsh realities of conflict. \\
\cline{1-4}
P17 & 1 & Quill & Viewers join friends in New Year's festivities, partaking in dances, concerts, and feasts, relishing the delight and warmth of shared experiences. \\
\cline{1-4}
P18 & 3 & Quill & Viewers explore a space blending Chinese zodiac themes, learning the importance of wisdom and persistence from the story of a monkey and a rooster.\\
\cline{1-4}
P19 & 0.5 & Quill & Viewers observe a farmer learning from his mistake of losing his sheep, mending the sheepfold in time, and understanding the importance of learning from errors. \\
\cline{1-4}
P20 & 12 & UE, Blender, Pro Tools & Viewers watch a kind sheep helping other cursed animals to sleep, thereby saving them, and in the end, discover themselves to be that sheep. \\
\cline{1-4}
P21 & 3 & Quill & Viewers observe a girl's fulfilling day in the old alleys behind skyscrapers, all under the watchful eyes of an extraterrestrial being living on the moon. \\ 
\bottomrule
\end{tabular}
\end{table*}

\end{document}